\documentclass[10pt,aps,pra,amssymb,superscriptaddress,footinbib,twocolumn]{revtex4-1}
\usepackage{graphicx}
\usepackage{amsmath}
\usepackage{amssymb}
\usepackage{color}
\usepackage{soul}
\usepackage{ulem} %to strike out
\usepackage{marginnote}
\usepackage{dsfont}
\usepackage{tikz}
\usepackage{hyperref}

\begin{document}

%\title{Entanglement, coherence, and measurement-induced conversion of coherence to entanglement in double spontaneous downconversion processes %bi-squeezed tripartite Gaussian states}

\title{Entanglement, coherence, and redistribution of quantum resources in double spontaneous downconversion processes}

\author{David Edward Bruschi}
\affiliation{York Centre for Quantum Technologies, Department of Physics, University of York, YO10 5DD Heslington, UK}
\author{Carlos Sab\'in}
\affiliation{Instituto de F\'isica Fundamental (CSIC),Serrano 113-bis, 28006 Madrid, Spain}
\author{Gheorghe Sorin Paraoanu}
\affiliation{Low Temperature Laboratory and Centre for Quantum Engineering, Department of Applied Physics, Aalto University School of Science, FI-00076 Aalto, Finland}

\begin{abstract}
We study the properties of bi-squeezed tripartite Gaussian states created by two spontaneous parametric down-conversion processes that share a common idler.
We give a complete description of the quantum correlations across of all partitions, as well as of the genuine multipartite entanglement, obtaining analytical expressions for most of the quantities of interest. We find that the state contains genuine tripartite entanglement, in addition to the bipartite entanglement among the modes that are directly squeezed. We also investigate the effect of homodyne detection of the photons in the common idler mode, and analyse the final reduced state of the remaining two signal modes.  We find that this measurement leads to a conversion of the coherence of the two signal modes into entanglement, a phenomenon that can be regarded as a redistribution of quantum resources between the modes. The applications of these results to quantum optics and circuit quantum electrodynamics platforms are also discussed.
\end{abstract}

\maketitle

%\tableofcontents

%------------------------------------------------------------------------------------------------------------------------------------------------------------------------%
\section{Introduction}
%------------------------------------------------------------------------------------------------------------------------------------------------------------------------%
The vacuum in quantum theory is one of the most subtle concepts in modern physics. The classical picture of a ``void'' or ``emptiness'' does not accurately capture the nature of this particular state, and new phenomena can be unveiled by systematically employing quantum mechanics \cite{Gea-Banacloche:Scully:1988}. In the language of quantum physics, the vacuum state is the lowest energy eigenstate of a particular field Hamiltonian.  One can picture the quantum vacuum as a state with some latent structure (see {\it e.g.} \cite{BostonStudies}), which can manifest, for example, through the conversion of quantum fluctuations into real excitations when some parameter in the Hamiltonian is changed (sudden quench, parametric driving, {\it etc.}). This phenomenon is generally known as dynamical Casimir effect \cite{moore}, as is typically exemplified by a mirror moving in vacuum at relativistic speeds \cite{Birrell:Davies:1984}. This effect can be demonstrated in the laboratory by using, for example, superconducting circuits \cite{Wilson:Johansson:2011,Paraoanu:Pnas}. In these scenarios, abrupt modifications of the boundary conditions \cite{Wilson:Johansson:2011} or of the speed of light in a meta-material \cite{Paraoanu:Pnas} by means of an external pump result in field excitations that can be amplified and detected. These processes give rise to two-mode squeezed microwaves which display entanglement \cite{Paraoanu:Pnas} and other forms of quantum correlations \cite{nonclassicaldce,discord,ipsteering}, triggering the question of their employability as resources for quantum technologies \cite{cluster,casimirsimone,Streltsov:Adesso:2016,Krenn:Hochrainer:2016}.

Recently, it has been reported that a new class of three-mode states can be generated in the laboratory by double-pumping a superconducting resonator \cite{Paraoanu:coherence}, with one mode common to both pumps. This can be regarded as a ``double dynamical Casimir effect'', since now the mirror moves under the action of two pumps with different frequencies - or, in other words, the motion of the mirror is a harmonic oscillation with the average frequency of the pumps modulated due to beating at half-difference frequency. One starts by considering three modes $a$,$b$ and $c$, where $b$ is the common mode (conventionally referred from now on as idler). The parametric processes are arranged such that one downconversion occurs between modes $a,b$ and the other between modes $b-c$. This clearly leads to two-mode squeezing between modes $a$ and $b$, and modes $b$ and $c$ respectively. However, we show that the resulting tripartite state not only does contain the standard correlations due to parametric two-mode squeezing, but it also displays coherence correlations among the modes $a$ and $c$, even if these modes are not directly connected by the pumps. The origin of this coherence is the lack of which-path information for the photons emitted in the idler.
The effect is analogous to the phenomenon of induced coherence without induced emission \cite{PhysRevLett.67.318,PhysRevA.44.4614}, with the difference that in quantum optics the parametric processes use the nonlinearity of an optical crystal, while in the case of the dynamical Casimir effect the system is linear and the pump changes an electrical/optical length or a boundary condition. In the following we will not distinguish between these two cases, as they are both instances of spontaneous parametric downconversion, that is, the decay of a pump photon into a signal photon and an idler photon, triggered by vacuum fluctuations.

Motivated by these experimental advances, we study here the entanglement and the coherence of bi-squeezed tripartite Gaussian states generated by double spontaneous parametric downconversions, deriving analytical results confirmed by numerical calculations. We provide a systematic analysis of the quantum correlation properties of the aforementioned class of tripartite Gaussian states. We find that there exists genuine tripartite entanglement above a threshold value of the initial squeezing parameter as well as $a$-$b$ and $b$-$c$ entanglement, but no $a$-$c$ entanglement.
We also propose an experiment, similar to postselection, where we perform homodyne detection on the common idler mode, and we calculate the covariance matrix for the remaining signal modes. Surprisingly, it appears that some of the coherence between modes $a$ and $c$ is converted into entanglement after the homodyne detection of mode $b$, providing an interesting example of redistribution of quantum resources.

We finally note that the study of tripartite systems has been first introduced in physics with the aim of understanding the fundamental quantum statistical properties of light and the nonlocal features of quantum physics, but interest in  applications for the development of quantum technologies has recently witnessed a resurgence.  The field is developing fast, with several experimental platforms being  used recently to generate tripartite states with high efficiency: cascaded parametric downconversion setups employing two \cite{Jennewein2010,PhysRevLett.109.253604} or three separate crystals \cite{PhysRevLett.113.173601}, nonlinear waveguides \cite{Krapick:16}, quantum dots \cite{Weihs}, and  hybrid systems comprising a Rb-85 hot atom cell and a nonlinear waveguide \cite{Ding:15}.
 The applications include for example quantum imaging \cite{Zeilingerimaging}, interferometry \cite{Lahiri2016,Hochrainer2016momentum,Hochrainer2016}, and quantum computing using networks and cluster states \cite{PhysRevLett.101.130501,PhysRevLett.107.030505,PhysRevLett.108.083601,Treps2014,PhysRevLett.112.120505}. The results presented here are device-independent, therefore they can be tested on any of these experimental platforms.

The paper is organised as follows. ln Section \ref{section:intro} we introduce the essential tools from quantum optics with continuous variables as well as the covariance matrix mathematical formalism.
The creation of bi-squeezed tripartite Gaussian states in systems driven parametrically by two pumps is then described in Section \ref{section:generation}.
In Section \ref{section:characterization} we study the genuine multipartite entanglement generated, the bipartite entanglement across all reduced states, and the coherence properties of these states.
Section \ref{section:meas} demonstrates the creation of entanglement between the signal modes under a homodyne measurement of the idler.
Next, in Section \ref{section:appl} we study a  few applications of these techniques to realistic scenarios, such as experiments at low temperatures or modes with very close frequencies. We discuss our results and the perspectives of this work in  a final conclusions section. For completeness, we provide details of the derivations in four appendices at the end of the paper.

We use in this work the following conventions: bold symbols stand for matrices and plain font with under scripts denote elements of vectors and matrices. In this work $\rm{Tp}$ stands for transposition, in order to avoid confusion with temperature, denoted by $T$, and time, denoted by $t$.

%------------------------------------------------------------------------------------------------------------------------------------------------------------------------%
\section{Continuous variables and Gaussian states}\label{section:intro}
%------------------------------------------------------------------------------------------------------------------------------------------------------------------------%
In this work we restrict our attention to Gaussian states of bosonic fields only. Gaussian states are a class of quantum states that enjoy remarkable properties, in particular when the transformations involved are linear unitary transformations, {\it i.e.} they are quadratic in the creation and annihilation operators \cite{RevModPhys.77.513}. In this case, the Gaussian character of the state is preserved and one can employ techniques from the covariance matrix formalism
\cite{Adesso:Ragy:2014}. Gaussian states of bosonic fields naturally occur in many experiments and, when applicable, offer a convenient description of the state of the electromagnetic field in the optical or microwave range. In this section we set the notations and we briefly introduce the main concepts of covariance matrix formalism, which is a powerful tool that can be used when considering unitary linear transformations between Gaussian states of bosonic fields.

%------------------------------------------------------------------------------------------------------------------------------------------------------------------------%
\subsection{Symplectic matrices}
%------------------------------------------------------------------------------------------------------------------------------------------------------------------------%

We start by considering $N$ bosonic modes ({\it e.g.}, harmonic oscillators) with annihilation and creation operators $a_n$ and $a^{\dag}_n$. These operators satisfy the standard canonical commutation relations $[a_n,a^{\dag}_{n'}]=\delta_{nn'}$, while all other commutators vanish. It is convenient to collect all the operators and introduce the vector $\mathbb{X}:=(a_1,a_2,...,a_N;a^{\dag}_1,a^{\dag}_2,...,a^{\dag}_N)^{\text{Tp}}$, where Tp stands for transpose. For example, we have $\mathbb{X}_2=a_2$ or $\mathbb{X}_{2N-1}=a^{\dag}_{N-1}$ with this choice of operator ordering.
We notice in passing that the techniques developed below can be extended in a straightforward fashion to an infinite number of bosonic operators. This situation occurs, for example, in quantum field theory in flat and curved spacetime \cite{Birrell:Davies:1984}.

The canonical commutation relations can now be written as $[\mathbb{X}_n,\mathbb{X}^{\dag}_m]=i\Omega_{nm}$, where $\Omega_{nm}$ are the elements of the $2N\times2N$ matrix $\boldsymbol{\Omega}$, known as symplectic form, which has the following expression
\begin{align}
i\boldsymbol{\Omega}=
\begin{pmatrix}
\mathds{1} & 0\\
0 & -\mathds{1}
\end{pmatrix}.
\end{align}
Here $\mathds{1}$ are the $N\times N$ identity matrices.

Any unitary transformation $U=\exp[-i\,H]$, with Hermitian generator $H$ quadratic in the creation and annihilation operators (or, equivalently, in the quadrature operators), induces a \textit{linear transformation} on the (collection of) operators $\mathbb{X}$  through the relation $U\,\mathbb{X}\,U^{\dag}=\boldsymbol{S}\,\mathbb{X}$.
The unitary operators in the expression $U\,\mathbb{X}\,U^{\dag}$ act on each element of the vector $\mathbb{X}$ independently and $\boldsymbol{S}$ is a symplectic matrix that takes the form $\boldsymbol{S}=\exp[-F\,\boldsymbol{\Omega}\,\boldsymbol{H}]$, where $F$ is a real function that needs to be determined, while $\boldsymbol{H}$ is defined through $H=\mathbb{X}^{\dag}\,\boldsymbol{H}\,\mathbb{X}$.

The matrix $\boldsymbol{S}$ is called symplectic since it satisfies $\boldsymbol{S}\,\boldsymbol{\Omega}\,\boldsymbol{S}^{\dag}=\boldsymbol{\Omega}$ or, equivalently, $\boldsymbol{S}^{\dag}\,\boldsymbol{\Omega}\,\boldsymbol{S}=\boldsymbol{\Omega}$. We note that $\text{det}(\boldsymbol{S})=1$, see {\it e.g.} \onlinecite{Arvind:1995}. 

A symplectic matrix $\boldsymbol{S}$ can always be written, with the particular choice of operator ordering in $\mathbb{X}$, as
\begin{align}
\boldsymbol{S}=
\begin{pmatrix}
\boldsymbol{\alpha} & \boldsymbol{\beta}\\
\boldsymbol{\beta}^* & \boldsymbol{\alpha}^*
\end{pmatrix}.
\end{align}
The $N\times N$ matrices $\boldsymbol{\alpha}$ and $\boldsymbol{\beta}$, in the case of quantum fields and curved spacetime, collect the well known Bogoliubov coefficients used extensively in literature \cite{Birrell:Davies:1984}.
These coefficients satisfy the well known Bogoliubov identities \cite{Birrell:Davies:1984}, which read $\boldsymbol{\alpha}\,\boldsymbol{\alpha}^{\dag}-\boldsymbol{\beta}\,\boldsymbol{\beta}^{\dag}=\mathds{1}$ and $\boldsymbol{\alpha}\,\boldsymbol{\beta}^{\rm Tp}-\boldsymbol{\beta}\,\boldsymbol{\alpha}^{\rm Tp}=0$ in compact form.

We finally notice that the formal machinery introduced here is \textit{independent} of the initial state of the system.

%------------------------------------------------------------------------------------------------------------------------------------------------------------------------%
\subsection{Gaussian states}
%------------------------------------------------------------------------------------------------------------------------------------------------------------------------%
Unitary evolution and transformations, represented by a unitary operator $U$, of bosonic systems which are initially in a state $\rho_i$ are of great importance in physics. Unitary evolution leads to a final state $\rho_f$ through the standard Heisenberg equation $\rho_f=U^{\dag}\,\rho_i\,U$.
If the state $\rho_i$ is a Gaussian state, and the unitary $U$ is a linear operator (see above), the Gaussian character is preserved; therefore, employing the specific results of Gaussian state formalism becomes very convenient \cite{Adesso:Ragy:2014}.

In general, a state $\rho$ of $N$ bosonic modes is defined by an infinite amount of degrees of freedom. However, a Gaussian state $\rho$ of bosonic modes is characterised only by a finite amount of degrees of freedom. In particular, it is uniquely defined by the vector $d$ of first moments and the second moments $\sigma_{nm}$ defined by $d:=\langle\mathbb{X}\rangle_{\rho}$ and $\sigma_{nm}:=\langle\{\mathbb{X}_n,\mathbb{X}^{\dag}_m\}\rangle_{\rho}-2\langle\mathbb{X}_n\rangle_{\rho}\langle\mathbb{X}^{\dag}_m\rangle_{\rho}$ respectively, see \cite{Adesso:Ragy:2014}.
Here, all expectation values $\langle \mathcal{O}\rangle_{\rho}$ of an operator $\mathcal{O}$ are defined by $\langle \mathcal{O}\rangle_{\rho}:=\text{Tr}(\mathcal{O}\rho)$ and $\{A,B\}=AB+BA$ is the anticommutator of operators $A$ and $B$.
In this work we ignore the first moments, which can be safely set to zero without loss of generality. We make this choice since we are interested in quantum correlations, which are unaffected by the first moments. Initial vanishing moments remain zero under symplectic transformations and the second moments $\sigma_{nm}$ can be conveniently collected into the Hermitian covariance matrix $\boldsymbol{\sigma}$.
We notice that a covariance matrix $\boldsymbol{\sigma}$ represents a physical state $\rho$ if it satisfies $\boldsymbol{\sigma}+i\,\boldsymbol{\Omega}\geq0$ in the operatorial sense \cite{Adesso:Ragy:2014}. This amounts to computing the usual eigenvalues of the matrix $\boldsymbol{\sigma}+i\,\boldsymbol{\Omega}$ and checking of they are positive.

We can now recast the Heisenberg equation $\rho_f=U^{\dag}\,\rho_i\,U$ into a relation between covariance matrices. Let the initial state $\rho_i$ be represented by the covariance matrix $\boldsymbol{\sigma}_i$ and the final state $\rho_f$ by the covariance matrix $\boldsymbol{\sigma}_f$. We have already seen that any quadratic unitary $U$ can be represented by a symplectic matrix $\boldsymbol{S}$. Then, the Heisenberg equation $\rho_f=U^{\dag}\,\rho_i\,U$ takes the form $\boldsymbol{\sigma}_f=\boldsymbol{S}^{\dag}\,\boldsymbol{\sigma}_i\,\boldsymbol{S}$, which reduces the problem of usually untreatable operator algebra to matrix multiplication of $2N\times2N$ matrices.

Williamson's theorem \cite{Williamson:1936:v1,Williamson:1936:v2,Arnold:1978} guarantees that any covariance matrix $\boldsymbol{\sigma}$ can be put in diagonal form by a symplectic matrix. This means that, given a covariance matrix $\boldsymbol{\sigma}$ it is always possible to find a symplectic matrix $\boldsymbol{s}$ such that $\boldsymbol{\sigma}=\boldsymbol{s}^{\dag}\,\boldsymbol{\nu}_{\oplus}\,\boldsymbol{s}$,
where the diagonal matrix $\boldsymbol{\nu}_{\oplus}=\text{diag}(\nu_1,\nu_2,...,\nu_N;\nu_1,\nu_2,...,\nu_N)$ is called the Williamson form of $\boldsymbol{\sigma}$ and $\nu_m\geq1$ are called the symplectic eigenvalues of $\boldsymbol{\sigma}$.
The symplectic eigenvalues $\{\pm\nu_m\}$ are obtained as the eigenvalues of the matrix $i\,\boldsymbol{\Omega}\,\boldsymbol{\sigma}$.
The purity $P$ of the state $\boldsymbol{\sigma}$ is given by $P=\prod_m\,\nu_m\geq1$, and the state is pure if $P=1$ (or, equivalently, $\nu_m=1$ for all $m$).

A $2N\times2N$ covariance matrix $\boldsymbol{\sigma}$ is a Hermitian matrix that can be written in the form
\begin{align}\label{general:covariance:matrix}
\boldsymbol{\sigma}=
\begin{pmatrix}
\boldsymbol{W} & \boldsymbol{V}\\
\boldsymbol{V}^* & \boldsymbol{W}^*
\end{pmatrix},
\end{align}
where the $N\times N$ matrices $\boldsymbol{W}$ and $\boldsymbol{V}$ satisfy $\boldsymbol{W}=\boldsymbol{W}^{\dag}$ and $\boldsymbol{V}=\boldsymbol{V}^{\rm{Tp}}$.

%------------------------------------------------------------------------------------------------------------------------------------------------------------------------%
\subsection{Useful properties of the covariance matrix}
%------------------------------------------------------------------------------------------------------------------------------------------------------------------------%
In this subsection we provide some useful insight on some properties enjoyed by the elements of the covariance matrices. We start by introducing the symplectic eigenvalues $\nu_m$. These eigenvalues can be written as $\nu_m=\coth\bigl(\frac{\hbar\,\omega_m}{2\,k_B\,T_n}\bigr)$, where $T_n$ is the \textit{local} temperature of the one-mode reduced state. The reason that the symplectic eigenvalues have this form results from the fact that every single-mode reduced state of a Gaussian state is a thermal state up to local operations \cite{Adesso:Ragy:2014}.
Notice that if a state $\boldsymbol{\sigma}$ is a thermal state then it coincides with its Williamson form, i.e., $\boldsymbol{\sigma}\equiv\boldsymbol{\nu}_{\oplus}$.

An important operation is the process of ``tracing out'' a particular subsystem. In this language, this operation just amounts to deleting the rows and columns corresponding to the system one wishes to trace out \cite{Adesso:Ragy:2014}.

As a useful application, we now show how we can employ the covariance matrix to compute quantities of interest. Let $\langle a^{\dag}_m\,a_m\rangle_{\rho}$ be the number expectation value of mode $m$. Without loss of generality, let us assume that the first moments vanish, i.e., $\langle\mathbb{X}\rangle=0$. Then it is easy to show that $\langle a^{\dag}_m\,a_m\rangle_{\rho}=\frac{1}{2}[\sigma_{mm}-1]$, which highlights the role of the covariance matrix when computing physically relevant quantities.

%------------------------------------------------------------------------------------------------------------------------------------------------------------------------%
\subsection{Entanglement in Gaussian states}
%------------------------------------------------------------------------------------------------------------------------------------------------------------------------%
The quantitative characterisation of entanglement is a central task in many areas of quantum science. For example, entanglement is at the core of quantum computation \cite{Lloyd:Braunstein:1999,Ladd:Jelezko:2010}, quantum cryptography and quantum communication \cite{Gisin:Ribordy:2002}. For two modes in general, and for Gaussian states in particular, the task has been fully solved in an unambiguous way, and two-mode entanglement has been completely characterised \cite{Adesso:Ragy:2014}.

It has been shown that every measure of entanglement for two mode symmetric Gaussian states is a function of the \textit{smallest symplectic eigenvalue} of the partial transpose \cite{Adesso:Ragy:2014}. This establishes the PPT criterion as the paramount criterion for two-mode symmetric Gaussian states, {\it i.e.} states for which the determinants of the reduced single modes states are the same.
One starts from the two mode state $\boldsymbol{\sigma}$ and obtains the partial transpose $\tilde{\boldsymbol{\sigma}}$ as $\tilde{\boldsymbol{\sigma}}=\boldsymbol{P}\,\boldsymbol{\sigma}\,\boldsymbol{P}$, where the partial transposition matrix $\boldsymbol{P}$ takes the form
\begin{align}
\boldsymbol{P}=
\begin{pmatrix}
1 & 0 & 0 & 0\\
0 & 0 & 0 & 1\\
0 & 0 & 1 & 0\\
0 & 1 & 0 & 0
\end{pmatrix}.
\end{align}
One then computes the symplectic eigenvalues $\{\tilde{\nu}_m\}$ of the partial transpose $\tilde{\boldsymbol{\sigma}}$ as the eigenvalues of the matrix $i\,\boldsymbol{\Omega}\,\tilde{\boldsymbol{\sigma}}$. These eigenvalues come in two pairs of identical eigenvalues and we denote the smallest one as $\tilde{\nu}_-$. If $\tilde{\nu}_-<1$ then there is entanglement.

The choice of a particular measure is a matter of convenience or of the problem at hand, since all measures are (decreasing) monotonic functions of $\tilde{\nu}_-$. We employ here the negativity $\mathcal{N}$ defined as
\begin{align}\label{negativity}
\mathcal{N}:=\text{max}\left[0,\frac{1-\tilde{\nu}_-}{2\,\tilde{\nu}_-}\right],
\end{align}
and the logarithmic negativity $E_{\mathcal{N}}$ defined as
\begin{align}\label{logarithmic:negativity}
E_{\mathcal{N}}:=\text{max}\left[0,-\ln(\tilde{\nu}_-)\right].
\end{align}
We can also choose the entanglement of formation for symmetric states $\mathcal{E}_{oF}$ defined as
\begin{align}\label{entanglement:of:formation}
\mathcal{E}_{oF}:=\text{max}\left[0,f_+(\tilde{\nu}_-)-f_-(\tilde{\nu}_-)\right],
\end{align}
where we have introduced the functions $f_{\pm}(x):=\frac{(x\pm1)^2}{4\,x}\ln\frac{(x\pm1)^2}{4\,x}$ for convenience of presentation.

%------------------------------------------------------------------------------------------------------------------------------------------------------------------------%
\subsection{Coherence in Gaussian states}
%------------------------------------------------------------------------------------------------------------------------------------------------------------------------%
The role of quantum coherence in emergent quantum technologies such as quantum thermodynamics, quantum metrology or quantum biology is currently the subject of intense research -- see the recent review \cite{Streltsov:Adesso:2016}, and so far there is no uniquely accepted measure of coherence.
Quantum coherence amounts to superposition with respect to a fixed orthonormal basis. A state is maximally incoherent (or mixed) if it is diagonal in the chosen basis. From here one can already see that the concept of coherence is linked to a choice of basis, therefore when using any measure of coherence one has to be specific. We choose to employ in the following two measures of coherence, a bipartite one defined operationally and based on interferometry, and a global one based on entropy.
The meaning of these measures is rather different: the first one refers only to two modes and characterises what occurs if these modes are combined by a beam-splitter. The second one measures how close is the state from a maximally mixed state, thus it provides a global measure of coherence that cumulates the information about all the possible correlations.

%------------------------------------------------------------------------------------------------------------------------------------------------------------------------%
\subsubsection{First-order bipartite quantum coherence}
%------------------------------------------------------------------------------------------------------------------------------------------------------------------------%
Given two modes $m$ and $n$, we call the correlation $\langle a_{m}^{\dag}\,a_{n}\rangle$
(first-order) bipartite coherence, sometimes denoted by $G^{(1)}_{mn}=\langle a^{\dag}_{m} a_{n}\rangle$ in optics
\cite{WallsMilburn}. This definition applies in general to any state, and therefore it can be used as well for Gaussian states.
This measures corresponds to a simple interferometric setup, where we collect the photons in the modes $m$ and $n$, add a phase difference between their paths, and let them interfere.   We will witness the formation of an interference pattern
only if the quantity $\langle a_{m}^{\dag}\,a_{n}\rangle$ is non-zero. This quantity can be normalized by the power in each mode, and in this case we recover the standard definition of first-order amplitude correlation function $g^{(1)}_{mn}$ from quantum optics applied to modes $m$ and $n$, namely
\begin{equation}
g^{(1)}_{mn} = \frac{\langle a^{\dag}_{m}\,a_{n}\rangle}{\sqrt{\langle a^{\dag}_{m}\,a_{m}\rangle\langle a^{\dag}_{n}\,a_{n}\rangle}}.
\end{equation}
Finally, we highlight a connection with many-body physics, where one often finds useful to employ the so-called single-particle density matrix $\boldsymbol{\rho^{(1)}}_{mn}$, see \cite{PenroseOsanger,BEC}, defined as
\begin{align}
\label{singlemodedensitymatrix}
\boldsymbol{\rho}^{\bf (1)}_{mn}=
\begin{pmatrix}
\langle a^{\dag}_{m}\,a_{m}\rangle & \langle a^{\dag}_{m}\,a_{n}\rangle \\
\langle a^{\dag}_{n}\,a_{m}\rangle & \langle a_{n}^{\dag}\,a_{n}\rangle \\
\end{pmatrix}.
\end{align}
The single-particle density matrix is an essential tool in the study of phase localization \cite{PhysRevA.55.4330,phaselocalization2008}
and fragmentation of Bose-Einstein condensates \cite{mueller,fragmentation2008} - where the vanishing of the off-diagonal element is used as a criterion for fragmentation (the single coherent wavefunction or order parameter associated with condensation breaks into a Fock state).

%------------------------------------------------------------------------------------------------------------------------------------------------------------------------%
\subsubsection{Relative entropy of coherence}
%------------------------------------------------------------------------------------------------------------------------------------------------------------------------%
A measure  of quantum coherence $C(\mu)$ for $N$-mode Gaussian states $\mu$ has been recently introduced \cite{xucohe} as $C(\mu)={\rm min}_{\delta}\{S(\mu||\delta)\}$, where $S(\mu||\delta) = {\rm Tr}[\mu \log_{2}\mu ] - {\rm Tr} [\mu \log_{2} \delta]$ is the relative entropy and
$\delta=\bigotimes_{k=1}^n\delta^k_{th}(\bar{n}_k)$ is a tensor product of reduced thermal states of each mode $k$. This measure is thus defined only in terms of the covariance matrix and displacement vectors.
The von Neumann entropy of the system in terms of the symplectic eigenvalues is given by:
\begin{align}
\label{entropy}
S(\mu)=&\sum_{k=1}^N\,\left[h_+(\nu_k)-h_-(\nu_k)\right],
\end{align}
where $h_{\pm}(x):=\left(\frac{x\pm 1}{2}\right)\ln\left(\frac{x\pm 1}{2}\right)$ and $\{\nu_k\}$ are the symplectic eigenvalues of $\mu$, while the mean occupation value is:
\begin{equation}\label{mean}
\bar{n}_k=\frac{1}{4}\left(\sigma^{(k)}_{11}+\sigma^{(k)}_{22}+[d_1^{(k)}]^2+[d_2^{(k)}]^2-2\right).
\end{equation}
Here $\sigma^{(k)}_{ij}$ and $[d_i^{(k)}]$ are the $ij$-th element of the reduced correlation matrix and the $i$ first statistical moment of the $k$ mode, respectively. In this work, the latter will always be equal to 0. It is possible to obtain an analytical expression in closed form \cite{xucohe}:
\begin{align}
\label{coherence}
C(\boldsymbol{\sigma})=&-S(\boldsymbol{\sigma})+\sum_{k=1}^n\left[(\bar{n}_k+1)\ln (\bar{n}_k+1)-\bar{n}_k\ln \bar{n}_k\right].
\end{align}

%------------------------------------------------------------------------------------------------------------------------------------------------------------------------%
\subsection{An example: Two-mode squeezing}
%------------------------------------------------------------------------------------------------------------------------------------------------------------------------%

To get a clear picture of the covariant matrix formalism, let us consider an useful example,
that of two-mode squeezing. In the experiments discussed further, two-mode squeezed states are produced by a single parametric process, {\it i.e.}
by the action of each pump acting separately.
 Let $\mathbb{X}:=(a,b,a^{\dag},b^{\dag})$. The unitary operator that implements two mode squeezing is $U(r)=\exp[r\,(a^{\dag}\,b^{\dag}-a\,b)]$ and it is easy to show that in this (simplified) case its symplectic representation is
\begin{align}
\boldsymbol{S}=
\begin{pmatrix}
\cosh r & 0 & 0 & \sinh r\\
0 & \cosh r & \sinh r & 0\\
0 & \sinh r & \cosh r & 0\\
\sinh r & 0 & 0 & \cosh r
\end{pmatrix}.\label{symplectic:two:mode:squeezing:matrix}
\end{align}
Notice that we have chosen a special case where the transformation is real, for the sake of simplicity and without any loss of generality. We can define the vector $\tilde{\mathbb{X}}:=(\tilde{a},\tilde{b},\tilde{a}^{\dag},\tilde{b}^{\dag})=\boldsymbol{S}\,\mathbb{X}$ of new operators. The two mode squeezing transformation reduces to its well-known form
\begin{align}
\tilde{a}=&\cosh r\, a+\sinh r\, b^{\dag} ,\nonumber\\
\tilde{b}=&\cosh r\, b+\sinh r\, a^{\dag}.\label{two:mode:squeezing:transformation}
\end{align}
In the usual Fock state formalism, the two-mode squeezed state $|\psi_r\rangle$ of two modes $a$ and $b$ has the form
\begin{align}\label{fock:state:two:mode:squeezed:state}
|\psi_r\rangle=\sum_{n=0}^{+\infty}\frac{\tanh^n r}{\cosh r}|n,n\rangle_{ab}.
\end{align}
In the covariance matrix formalism, we can easily see that the two-mode squeezed state \eqref{fock:state:two:mode:squeezed:state} takes the form
\begin{align}\label{two:mode:squeezed:state}
\boldsymbol{\sigma}=
\begin{pmatrix}
\cosh 2r & 0 & 0 & \sinh 2r\\
0 & \cosh 2r & \sinh 2r & 0\\
0 & \sinh 2r & \cosh 2r & 0\\
\sinh 2r & 0 & 0 & \cosh 2r
\end{pmatrix}.
\end{align}
This explicitly shows how Gaussian states in the Fock state formalism reduce to simple matrices in the Gaussian state formalism. In particular, simple analytical formulae are known for calculating fidelities \cite{PhysRevA.61.022306} and distance measures \cite{PhysRevA.58.869}.

We now proceed and compute the spectrum $\{\tilde{\nu}_m\}$ of the matrix $i\,\boldsymbol{\Omega}\,\boldsymbol{P}\,\boldsymbol{\sigma}\,\boldsymbol{P}$, which is the set of the symplectic eigenvalues of the partial transpose of the state \eqref{two:mode:squeezed:state}.
It is easy to show that they are $\{\tilde{\nu}_m\}=(e^{2\,r},e^{2\,r},e^{-2\,r},e^{-2\,r})$. We see that the smallest symplectic eigenvalue $\tilde{\nu}_-$ has the expression $\tilde{\nu}_-=e^{-2\,r}$. This implies that the logarithimic negativity $E_{\mathcal{N}}$ reads $E_{\mathcal{N}}=2\,r$, see \cite{Ohliger2010}.
The coherence for two-mode squeezed states can be calculated as well. In the case of the interference-based bipartite coherence, we find that $\langle \psi_r| a^{\dag} b |\psi_r \rangle =0$, which is a consequence of the peculiar structure of the state in the number basis Eq. (\ref{fock:state:two:mode:squeezed:state}). The entropy of coherence however gives a non-zero result which grows monotonically with $r$:
\begin{equation}
C(\boldsymbol{\sigma})=4\left[\cosh^2r\,\log_2\cosh r-\sinh^2r\,\log_2\sinh r\right]
\end{equation}

Once more, this underlines the power of the covariance matrix formalism, where simple analytical expressions can be obtained for the relevant quantities.

%------------------------------------------------------------------------------------------------------------------------------------------------------------------------%
\section{Generation of bi-squeezed tri-partite Gaussian states}\label{section:generation}
\begin{figure}[h!]
\includegraphics[width=0.5\textwidth]{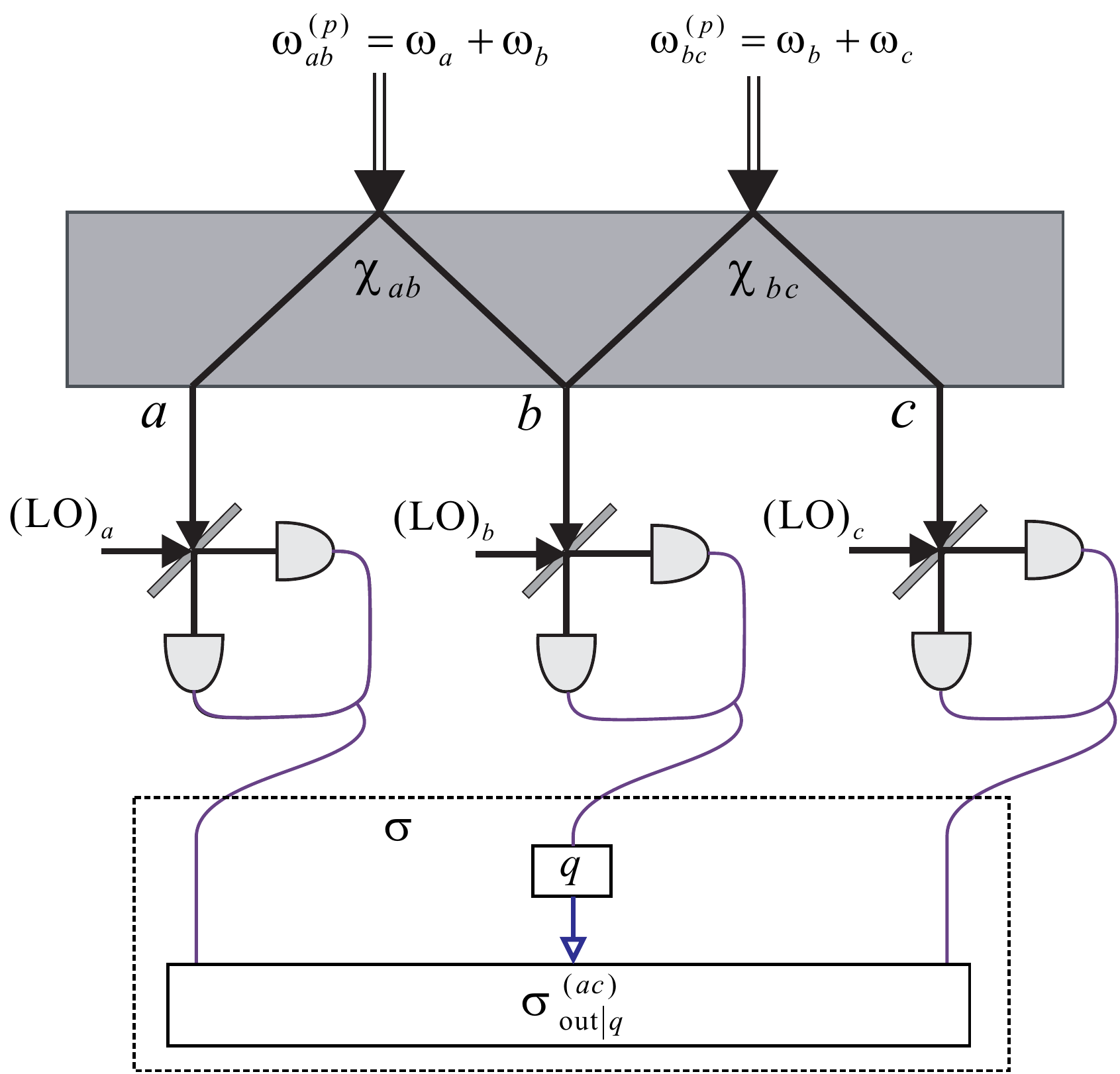}
\caption{Generic representation of the system of interest and the mesurement configuration. We use a semi-abstract representation for the parametric generation of photons in the three modes $a$, $b$, and $c$. This can be realized using either one resonator with mode $b$ as idler common to the two pumps $\omega_{ab}^{(p)}$ and $\omega_{bc}^{(p)}$ or two different nonlinear crystals with idlers aligned with each other. Each mode can be measured by a homodyne detection scheme using oscillators $(\mathrm{LO})_a$,  $(\mathrm{LO})_b$, $(\mathrm{LO})_c$. One can either measure the 3x3 correlation matrix $\sigma$ or one can perform only a $q$-quadrature measurement in mode $b$, to be left with a state ${\bf \sigma}^{(ac)}_{{\rm out}|q}$.}
\label{fig:fig1}
\end{figure}

%------------------------------------------------------------------------------------------------------------------------------------------------------------------------%
We now move to the physical system of interest, see Fig. \ref{fig:fig1}. This consists of three bosonic modes $a,b,c$ with frequencies $\omega_a,\omega_b,\omega_c$ respectively. The three modes are modulated parametrically by two pump fields at the frequencies $\omega_{ab}^{(p)}$ and $\omega_{bc}^{(p)}$. Systems of this type have been studied experimentally, both in the optical and in the microwave frequency range. To encompass all the physical realizations, we use a semi-abstract, device-independent representation \cite{Krenn:Hochrainer:2016} which shows the mode $b$ as common to two parametric processes $(ab)$ and $(bc)$ occurring in a parametrically device pumped at
$\omega_{ab}^{(p)}$ and $\omega_{bc}^{(p)}$. In practice, this can be realized by overlapping the paths of the idler photons of two different optical crystals, by using a single nonlinear crystal in a multimode cavity ({\it e.g.} bismuth borate (BIBO) in a ring cavity \cite{Treps2014}, periodically poled
KTiOPO$_4$(PPKTP) with zzz quasi-phase-
matching \cite{PhysRevLett.112.120505,PhysRevLett.107.030505}), or a single superconducting resonator with double-modulated electrical length \cite{Paraoanu:coherence}.
The total Hamiltonian $H$ for this type of configuration can be constructed by adding to the harmonic-oscillator Hamiltonian $H_{0} = \hbar\,\omega_{a}\,a^{\dag}\,a +
\hbar\,\omega_{b}\,b^{\dag}\,b + \hbar\,\omega_{c}\,c^{\dag}c$ of the three modes, two parametric perturbations corresponding to each pump \cite{WallsMilburn}. One obtains the total Hamiltonian $H$ in the form
\begin{equation}
H = H_0 + \hbar\,\left( \chi^{*}_{ab}\,e^{i \omega_{ab}^{(p)}t}\,a\,b + \chi^{*}_{bc}\,e^{i \omega_{bc}^{(p)}t}\,b\,c + {\rm h.c.} \right). \label{hamiltonian:generic}
\end{equation}
Here $\chi_{ab}$ and $\chi_{bc}$ have dimensions of frequencies and describe the parametric coupling of the pumping fields into the modes $a,b,c$, representing the parametric analog of the Rabi frequency of driven two-level systems. This prescription is very general, irrespective to the particular physical system employed or to weather the modulation is done on the boundary conditions or in the
bulk of the material or device (see {\it e.g.} Supporting Information in \cite{Paraoanu:Pnas} and \cite{goran:pra}).
Consider now the unitary transformation $U_{0}(t) = \exp[- \frac{i}{\hbar}\,H_{0}\,t]$. Assuming that the frequencies of the pumps $\omega_{ab}^{(p)}$ and $\omega_{bc}^{(p)}$ are chosen such that the energy conservation conditions $\omega_{ab}^{(p)}=\omega_a+\omega_b$ and $\omega_{bc}^{(p)}=\omega_b+\omega_c$ are satisfied, the Hamiltonian \eqref{hamiltonian:generic} can be transformed into $H_{\rm eff} = U_{0}(t)^{\dag} H U_{0}(t) + i \hbar[dU_{0}^{\dag}(t)/dt]U_{0}(t)$,
where we have defined
\begin{equation}
H_{\rm eff} = \hbar \left( \chi^{*}_{ab}\,a\,b + \chi^{*}_{bc}\,b\,c + {\rm h.c.} \right).
\end{equation}
The effective Hamiltonian $H_{\rm eff}$ is now time-independent and describes the evolution of the system in a triple rotating frame (with frequencies $\omega_a,\omega_b,\omega_c$).
Suppose now that the system is pumped for a finite time $\tau$, as it was done in the time-domain experiments in previous work \cite{Paraoanu:coherence}. Then, introducing the two mode squeezing operators $G_{ab}:=a^{\dag}\,b^{\dag}+a\,b$ and $G_{bc}:=b^{\dag}\,c^{\dag}+b\,c$
and the corresponding two-mode squeezing parameters $R_{ab} = -\chi_{ab}^{*}\,\tau$ and $R_{bc} = -\chi_{bc}^{*}\,\tau$
we get
\begin{align}\label{sorin:operator:first}
U=e^{i\,\left[R_{ab}\,G_{ab}+R_{bc}\,G_{bc}\right]} 
\end{align}
Here we implicitly assume that the resonators or cavities have a high enough quality factor, ensuring that absolute values of the parametric coupling is larger than the decay rate. Also,  after preparation, the measurement is realized on a timescale smaller than the relaxation time. These conditions are easily met in the present optical or superconducting-circuits setups. For example, in the experiments realized with a SQUID-based modulated resonator with decay rate of 1 MHz \cite{Paraoanu:coherence} the correlations were measured in time-domain, under the double parametric excitation of the system with 1 $\mu$s microwave pulses. Thus, for this system the conditions above are easily satisfies for $0 <\tau \ll 1\mu$s. These results can be readily extended to larger timescales and the signal can be enhanced with the use of higher-Q resonators. Moreover, we emphasize that the same structure comprising two two-mode squeezing operators can be recovered also in frequency space for continuous pumping of systems with dissipation in the input-output formalism \cite{Paraoanu:coherence}.

To investigate systematically the correlations induced by this operator, we collect the creation and annihilation operators of these modes in the vector $\mathbb{X}=(a,b,c,a^{\dag},b^{\dag},c^{\dag})^{\rm Tp}$.
We assume that the initial state is a thermal state $\boldsymbol{\sigma}_{\rm th}$ at temperature $T$, since temperature is always present in any real system. As mentioned before, in this case the state $\boldsymbol{\sigma}_{\rm th}$ of the system coincides with its Williamson form, i.e., $\boldsymbol{\sigma}_{\rm th}=\boldsymbol{\nu}_{\oplus}=\text{diag}(\nu_a,\nu_b,\nu_c,\nu_a,\nu_b,\nu_c)$.

Next, we proceed to construct the final state of interest $\rho$, represented by the covariance matrix $\boldsymbol{\sigma}$, that we obtain by applying the operator $U$
to the thermal state $\boldsymbol{\sigma}_{th}$.
For simplicity, we assume that the squeezing parameters $R_{ab}$ and $R_{bc}$ are both real. This is not a loss of generality: indeed, if the pumps have nonzero phases, 
$\chi_{ab}= |\chi_{ab}|e^{i\varphi_{ab}}$, $\chi_{bc}= |\chi_{bc}|e^{i\varphi_{bc}}$ we can obtain the evolution $U$ from Eq. \ref{sorin:operator:first} with real $R_{ab} = -|\chi_{ab}|\tau $ and $R_{bc} = -|\chi_{bc}|\tau$ by redefining 
$a\rightarrow a e^{i\theta_{a}}$, $b\rightarrow b e^{i\theta_{b}}$, $c\rightarrow c e^{i\theta_{c}}$ such that $\theta_{a}+ \theta_{b} = \varphi_{ab}$ and $\theta_{b}+ \theta_{c} = \varphi_{bc}$.

In general, it is possible to compute the state $\rho$ the symplectic matrix $\boldsymbol{S}$ representing the operator \eqref{sorin:operator:first} in the Fock state formalism. However, the results can be extremely difficult to manage analytically.

Here we use a recently-developed technique \cite{Bruschi:Lee:2013,Moore:Bruschi:2016} (see also \cite{Brown:MartinMartinez:2013} for an alternative approach) to obtain a more convenient representation of the operator \eqref{sorin:operator:first}, based on the Lie algebra structure of the $SL(3,\mathbb{C})$ group \cite{Arnold:1978}. In Appendix \ref{appendix:zero} we show that it is possible to re-write the operator \eqref{sorin:operator:first} as
\begin{align}\label{decoupled:representation}
U=e^{i\,\theta_{ac}\,B_{ac}}\,e^{i\,r_{ab}\,G_{ab}}\,e^{i\,r_{bc}\,G_{bc}},
\end{align}
where the real squeezing parameters $r_{ab},r_{bc}$ and phase $\theta_{ac}$ have the \textit{exact} expression
\begin{align}\label{final:result:first}
r_{ab}=&\ln\left(\cos\phi\,\sinh\rho+\sqrt{1+\cos^2\phi\,\sinh^2\rho}\right),\nonumber\\
r_{bc}=&\frac{1}{2}\ln\left(\frac{1+\sin\phi\,\tanh\rho}{1-\sin\phi\,\tanh\rho}\right),\nonumber\\
\theta_{ac}=&\arctan\left(\frac{\tan\phi}{\cosh\rho}\right)-\phi ,
\end{align}
as functions of the new parameters $\rho:=\sqrt{R_{ab}^2+R_{bc}^2}$ and $\tan \phi:=R_{bc}/R_{ab}$. Here $B_{ac}=i\,\left[a\,c^{\dag}-c\,a^{\dag}\right]$ is a beam-splitter transformation. The result is remarkable, because in general it is not possible to obtain simple analytical solutions when trying to factorize an exponential of multiple-mode operators using the well-known Hausdorff-Baker-Campbell approach to decoupling exponentials. We emphasise that the unitary operators \eqref{sorin:operator:first} and \eqref{decoupled:representation} are equivalent, and the final state obtained under their action is also the same.
%The choice of  \eqref{sorin:operator:first} or \eqref{decoupled:representation} is a matter of experimental or theoretical convenience.

We note that an alternative technique to decouple Eq. \eqref{sorin:operator:first} has been developed \cite{Braunstein:2005}. This yields a global passive operation, followed by a set of single mode squeezers, followed by another global passive transformation. Differently from this, the decomposition \eqref{decoupled:representation} comprises a series of two mode squeezers, which also gives a direct operational meaning as a sequence achievable in experiments. Specifically, the factorized representation \eqref{decoupled:representation} can be used as well as a heuristic tool in designing novel experiments, since the bi-squeezed Gaussian state obtained by double parametric pumping can be created also by pumping first one pair of modes, then another pair, and finally performing a beam-splitter transformation. For example, in quantum optics it might be convenient to even use two different crystals for realizing the two squeezing operations.

In order to obtain the correlation matrix, we start by applying a beam splitting $\boldsymbol{S}_{ac}(\theta_{ac})$  on modes $a$ and $c$ which, in symplectic geometry, has the form
\begin{align}\label{beam:splitting:modes:ac}
\boldsymbol{S}_{ac}(\theta_{ac})=
\begin{pmatrix}
\cos\theta_{ac} & 0 & \sin\theta_{ac} & 0 & 0 & 0\\
0 & 1 & 0 & 0 & 0 & 0\\
-\sin\theta_{ac} & 0 & \cos\theta_{ac} & 0 & 0 & 0\\
0 & 0 & 0 & \cos\theta_{ac} & 0 & \sin\theta_{ac}\\
0 & 0 & 0 & 0 &1 & 0\\
0 & 0 & 0 & -\sin\theta_{ac} & 0 & \cos\theta_{ac}\\
\end{pmatrix}.
\end{align}
Notice that this would be a trivial operation if the state was the vacuum; however, the initial state is thermal and the beamsplitter can have a non-trivial effect.

We proceed by applying a two mode squeezing $\boldsymbol{S}_{ab}(r_{ab})$ on modes $a$ and $b$ and a two mode squeezing $\boldsymbol{S}_{bc}(r_{bc})$  on modes $b$ and $c$.
These have the form
\begin{align}\label{two:mode:squeezing:modes:ab}
\boldsymbol{S}_{ab}(r_{ab})=
\begin{pmatrix}
{\rm ch}_{ab} & 0 & 0 & 0 & {\rm sh}_{ab} & 0\\
0 & {\rm ch}_{ab} & 0 & {\rm sh}_{ab} & 0 & 0\\
0 & 0 & 1 & 0 & 0 & 0\\
0 & {\rm sh}_{ab} & 0 & {\rm ch}_{ab} & 0 & 0\\
{\rm sh}_{ab} & 0 & 0 & 0 & {\rm ch}_{ab} & 0\\
0 & 0 & 0 & 0 & 0 & 1\\
\end{pmatrix}.
\end{align}
and
\begin{align}\label{two:mode:squeezing:modes:bc}
\boldsymbol{S}_{bc}(r_{bc})=
\begin{pmatrix}
1 & 0 & 0 & 0 & 0 & 0\\
0 & {\rm ch}_{bc} & 0 & 0 & 0 & {\rm sh}_{bc}\\
0 & 0 & {\rm ch}_{bc} & 0 & {\rm sh}_{bc} & 0\\
0 & 0 & 0 & 1 & 0 & 0\\
0 & 0 & {\rm sh}_{bc} & 0 & {\rm ch}_{bc} & 0\\
0 & {\rm sh}_{bc} & 0 & 0 & 0 & {\rm ch}_{bc}\\
\end{pmatrix},
\end{align}
where we have introduced ${\rm ch}_{ab}:=\cosh r_{ab},{\rm sh}_{ab}:=\sinh r_{ab}$, ${\rm th}_{ab}:=\tanh r_{ab}$ and ${\rm ch}_{bc}:=\cosh r_{bc},{\rm sh}_{bc}:=\sinh r_{bc}$, ${\rm th}_{bc}:=\tanh r_{bc}$ for compactness and simplicity of presentation.

The final state, when acting on the thermal state $\boldsymbol{\sigma}_{th}=\text{diag}(\nu_a,\nu_b,\nu_c,\nu_a,\nu_b,\nu_c)$ as well as all the reduced states, can be obtained analytically, see Appendix \ref{appendix:one} for the full expressions of each matrix element. Here we report only the structure of these states, which is essential for the ensuing calculations. The three-mode state $\boldsymbol{\sigma}$ reads
\begin{align}\label{final:three:mode:state:structure}
\boldsymbol{\sigma}=
\begin{pmatrix}
\alpha & 0 & \delta & 0 & \epsilon & 0\\
0 & \beta & 0 & \epsilon & 0 & \zeta\\
\delta & 0 & \gamma & 0 & \zeta & 0\\
0 & \epsilon & 0 & \alpha & 0 & \delta\\
\epsilon & 0 & \zeta & 0 & \beta & 0\\
0 & \zeta & 0 & \delta & 0 & \gamma\\
\end{pmatrix}.
\end{align}
The final two-mode $\boldsymbol{\sigma}^{(ab)},\boldsymbol{\sigma}^{(bc)},\boldsymbol{\sigma}^{(ac)}$ and single-mode states $\boldsymbol{\sigma}^{(a)},\boldsymbol{\sigma}^{(b)},\boldsymbol{\sigma}^{(c)}$ read
\begin{align}\label{final:two:mode:singl:mode:state:structure}
\boldsymbol{\sigma}^{(ab)}=
\begin{pmatrix}
\alpha & 0 & 0 & \epsilon \\
0 & \beta & \epsilon & 0 \\
0 & \epsilon & \alpha & 0 \\
\epsilon & 0 & 0 & \beta
\end{pmatrix}
&
\,\,\,\,\,\,
\boldsymbol{\sigma}^{(a)}=
\begin{pmatrix}
\alpha & 0 \\
0 & \alpha
\end{pmatrix},
\nonumber\\
\boldsymbol{\sigma}^{(bc)}=
\begin{pmatrix}
\beta & 0 & 0 & \zeta \\
0 & \gamma & \zeta & 0 \\
0 & \zeta & \beta & 0 \\
\zeta & 0 & 0 & \gamma
\end{pmatrix}
&
\,\,\,\,\,\,
\boldsymbol{\sigma}^{(b)}=
\begin{pmatrix}
\beta & 0 \\
0 & \beta
\end{pmatrix},
\nonumber\\
\boldsymbol{\sigma}^{(ac)}=
\begin{pmatrix}
\alpha & \delta & 0 & 0 \\
\delta & \gamma & 0 & 0 \\
0 & 0 & \alpha & \delta \\
0 & 0 & \delta & \gamma
\end{pmatrix}
&
\,\,\,\,\,\,
\boldsymbol{\sigma}^{(c)}=
\begin{pmatrix}
\gamma & 0 \\
0 & \gamma
\end{pmatrix}.
\end{align}
The reduced two-modes and single-mode covariance matrices were obtained using the \textit{trace-out} prescription from Section II C, namely eliminating one, and respectively two, modes from the three-mode matrix \eqref{final:three:mode:state:structure}. Finally, notice that all reduced single mode states are thermal.

%------------------------------------------------------------------------------------------------------------------------------------------------------------------------%
\section{Characterising bi-squeezed tripartite Gaussian states} \label{section:characterization}
%------------------------------------------------------------------------------------------------------------------------------------------------------------------------%
In this section we present a full description of the bi-squeezed tripartite Gaussian states create by the double parametric pumping described in the previous section. In particular, we focus on the entanglement properties, showing that the state has so-called genuine tripartite entanglement, and on the phenomenon of induced coherence between
modes $a$ and $c$ due to the indistinguishability of the photons in the common idler $b$.

%------------------------------------------------------------------------------------------------------------------------------------------------------------------------%
\subsection{Number expectation values}
%------------------------------------------------------------------------------------------------------------------------------------------------------------------------%
We can now turn to computing the final number expectation value for all three modes.
We have
\begin{align}\label{number:expectation:value:equations}
\langle a^{\dag}\,a\rangle=&\frac{1}{2}\left[\alpha-1\right],\nonumber\\
\langle b^{\dag}\,b\rangle=&\frac{1}{2}\left[\beta-1\right], \nonumber\\
\langle c^{\dag}\,c\rangle=&\frac{1}{2}\left[\gamma-1\right].
\end{align}

%------------------------------------------------------------------------------------------------------------------------------------------------------------------------%
\subsection{Purity of all reduced states}
%------------------------------------------------------------------------------------------------------------------------------------------------------------------------%
We wish to understand the correlation structure of the whole state. A rough understanding can be already given by computing the purity of all the reduced states.

Let us start with the purity $P_{\rm th}=\nu_a^2\,\nu_b^2\,\nu_c^2$ of the initial global tri-partite thermal state, which remains unchanged under our unitary transformations. We then list the initial purities of the thermal state $P_{\rm th}^{ab},P_{\rm th}^{bc},P_{\rm th}^{ac},P_{\rm th}^{a},P_{\rm th}^{b},P_{\rm th}^{c}$ which read
\begin{align}\label{initial:purities}
P_{\rm th}^{(ab)}=&\nu_a^2\,\nu_b^2\,\,\,\,\,\,\,\,&P_{\rm th}^{(a)}=&\nu_a^2 ,\nonumber\\
P_{\rm th}^{(bc)}=&\nu_b^2\,\nu_c^2\,\,\,\,\,\,\,\,&P_{\rm th}^{(b)}=&\nu_b^2 , \nonumber\\
P_{\rm th}^{(ac)}=&\nu_a^2\,\nu_c^2\,\,\,\,\,\,\,\,&P_{\rm th}^{(c)}=&\nu_c^2.
\end{align}
We now find that the purities of all reduced states of our given state are
\begin{align}\label{final:purities}
P^{(ab)}=&(\alpha\,\beta-\epsilon^2)^2\,\,\,\,\,\,\,\,&P^{(a)}=\alpha^2 , \nonumber\\
P^{(bc)}=&(\beta\,\gamma-\zeta^2)^2\,\,\,\,\,\,\,\,&P^{(b)}=\beta^2 , \nonumber\\
P^{(ac)}=&(\alpha\,\gamma-\delta^2)^2\,\,\,\,\,\,\,\,&P^{(c)}=\gamma^2.
\end{align}
We see that local purities have changed from the values in (\ref{initial:purities}) to the ones in (\ref{final:purities}), therefore we expect some correlations between the different modes. We proceed to study this in the next section.

%------------------------------------------------------------------------------------------------------------------------------------------------------------------------%
\subsection{Tripartite entanglement}
%------------------------------------------------------------------------------------------------------------------------------------------------------------------------%
Here we look at the nature of (quantum) correlations in the tripartite state of interest in this work. We  study the global (genuine) correlations as well as the bipartite correlations across all bipartite reduced states.

\begin{figure}[h!]
\includegraphics[width=0.5\textwidth]{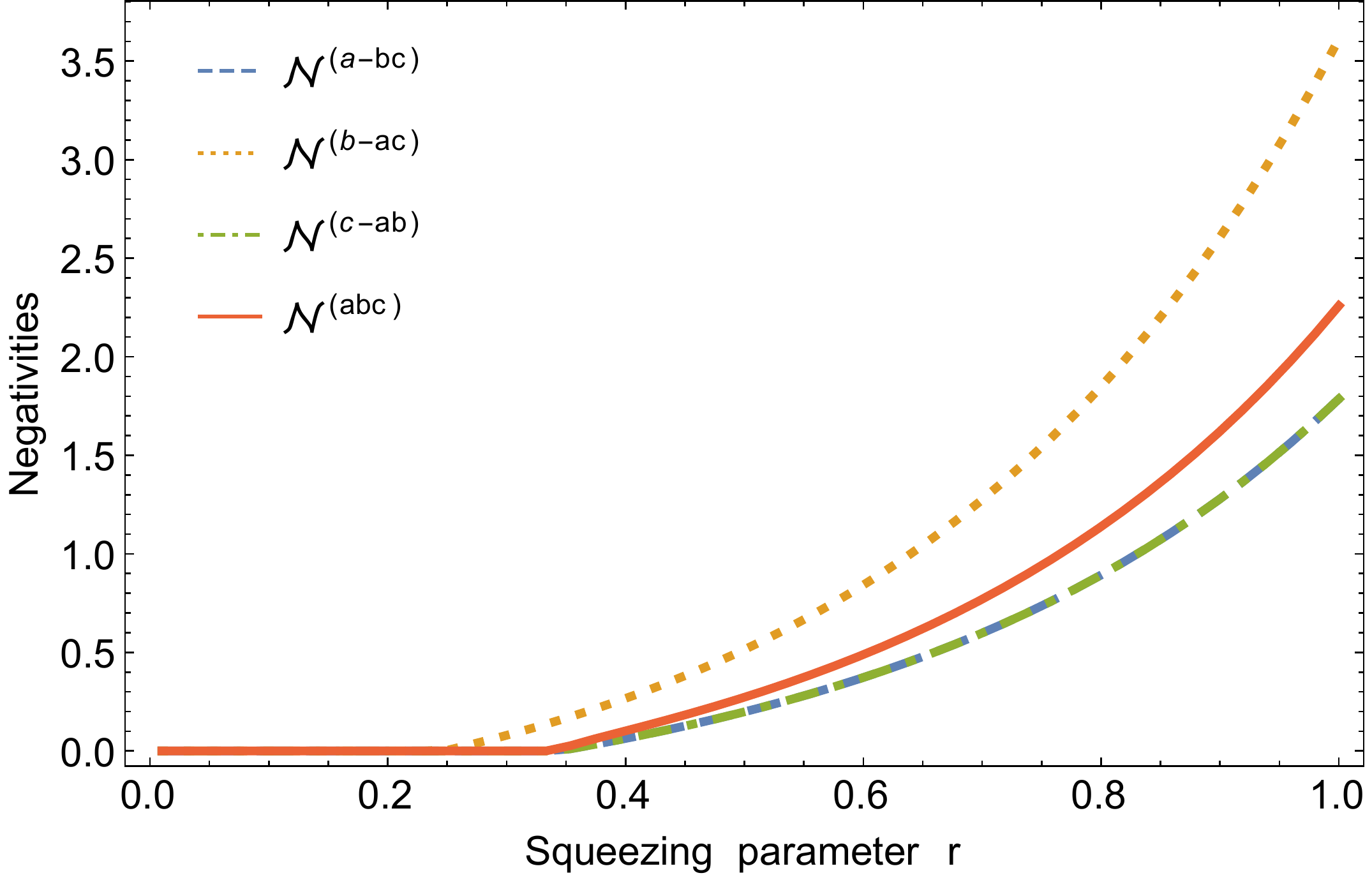}
\caption{
Tripartite negativity $\mathcal{N}^{(abc)}$, $\mathcal{N}^{(a-bc)}$, $\mathcal{N}^{(b-ac)}$ and $\mathcal{N}^{(c-ab)}$  vs. squeezing parameter $r=R_{ab}=R_{bc}$ for $\omega_a = 2\pi \times 4.99\operatorname{GHz}$, $\omega_b =2\pi \times 5\operatorname{GHz}$, $\omega_c = 2\pi \times 5.01\operatorname{GHz}$ and $T=15\operatorname{mK}$.}
\label{fig:fig2}
\end{figure}

A measure of the tripartite entanglement can be obtained through a suitable average of the entanglement of all the bipartitions of the system. For instance, we can consider the tripartite negativity $\mathcal{N}^{(abc)}$ defined by
\begin{equation}\label{eq:trineg}
\mathcal{N}^{(abc)}=\left[\mathcal{N}^{(a-bc)}\,\mathcal{N}^{(b-ac)}\,\mathcal{N}^{(c-ab)}\right]^{\frac{1}{3}},
\end{equation}
where $\mathcal{N}^{(i-jk)}$  is the negativity of the $i-jk$ bipartition as provided by the partial transposition with respect to the mode $i$.
In Fig. \ref{fig:fig2} we plot all the $\mathcal{N}^{(i-jk)}$ and the resulting $\mathcal{N}^{(abc)}$.

We notice that there is need for a certain amount of squeezing before genuine multipartite entanglement can be created.

%------------------------------------------------------------------------------------------------------------------------------------------------------------------------%
\subsection{Bipartite entanglement and coherence}
%------------------------------------------------------------------------------------------------------------------------------------------------------------------------%

%------------------------------------------------------------------------------------------------------------------------------------------------------------------------%
\subsubsection{Bipartite entanglement in the ``$ab$'' and ``$bc$'' subsystems}
%------------------------------------------------------------------------------------------------------------------------------------------------------------------------%
As previously discussed, we now need to compute the smallest symplectic eigenvalue $\tilde{\nu}_-$ for each reduced state. This eigenvalue provide us with a quantification of entanglement.

We start by the reduced state $\boldsymbol{\sigma}^{(ab)}$ of modes $a$ and $b$. We can compute the smallest symplectic eigenvalue $\tilde{\nu}_-^{(ab)}$ of the partial transpose and we find
\begin{align}
\tilde{\nu}_-^{(ab)}=\frac{1}{2}\left[\alpha+\beta-\sqrt{(\alpha-\beta)^2+4\,\epsilon^2}\right].
\end{align}
Similarly, for the reduced state $\boldsymbol{\sigma}^{(bc)}$ of modes $b$ and $c$ we can compute the smallest symplectic eigenvalue $\tilde{\nu}_-^{(bc)}$ of the partial transpose,
\begin{align}
\tilde{\nu}_-^{(bc)}=\frac{1}{2}\left[\beta+\gamma-\sqrt{(\beta-\gamma)^2+4\,\zeta^2}\right].
\end{align}
These eigenvalue can now be used, together with Eqs. \eqref{negativity} and \eqref{entanglement:of:formation}, to compute the negativities $\mathcal{N}^{(ab)}$, $\mathcal{N}^{(bc)}$ as well as the entanglement of formation $\mathcal{E}_{oF}^{(ab)}$, $\mathcal{E}_{oF}^{(bc)}$ in the reduced states $\boldsymbol{\sigma}^{(ab)}$ respectively $\boldsymbol{\sigma}^{(bc)}$.

%------------------------------------------------------------------------------------------------------------------------------------------------------------------------%
\subsubsection{Bipartite entanglement in the ``$ac$'' subsystem}
%------------------------------------------------------------------------------------------------------------------------------------------------------------------------%
Next, we calculate the reduced state $\boldsymbol{\sigma}^{(ac)}$ of modes $a$ and $c$. We can compute the smallest symplectic eigenvalue $\tilde{\nu}_-^{(ac)}$ of the partial transpose and we find
\begin{align}
\tilde{\nu}_-^{(ac)}=\frac{1}{2}\left[\sqrt{(\alpha+\gamma)^2-4\,\delta^2}-|\alpha-\gamma|\right].
\end{align}
It is easy to show that $\tilde{\nu}_-^{(ac)}\geq2\,|\alpha\gamma-\delta^2|$. This, in turn can be written as $\tilde{\nu}_-^{(ac)}\geq2\,\sqrt{P^{(ac)}}$, where $P^{(ac)}$ is the purity of the final reduced state. Since the purity $P$ of \textit{any} state, in this language, satisfies $P\geq1$ we conclude that
\begin{align}
\tilde{\nu}_-^{(ac)}\geq2,
\end{align}
which implies that there can never be any entanglement between the modes $a$ and $c$, as expected.

The bipartite negativities $\mathcal{N}^{(ij)}$ between modes i and j, where $i,j \in \{a,b,c\}$ are plotted in Fig. (\ref{fig:fig3}). We observe that while $\mathcal{N}^{(bc)}=\mathcal{N}^{(ab)}$ is different from 0 and grows with the initial squeezing, the negativity $\mathcal{N}^{(ac)}$ is 0 for any value of the initial squeezing - as expected. Next, we proceed to study the issue of coherence.

%------------------------------------------------------------------------------------------------------------------------------------------------------------------------%
\subsubsection{Bipartite coherence and relative entropy of coherence}
\begin{figure}[h!]
\includegraphics[width=0.5\textwidth]{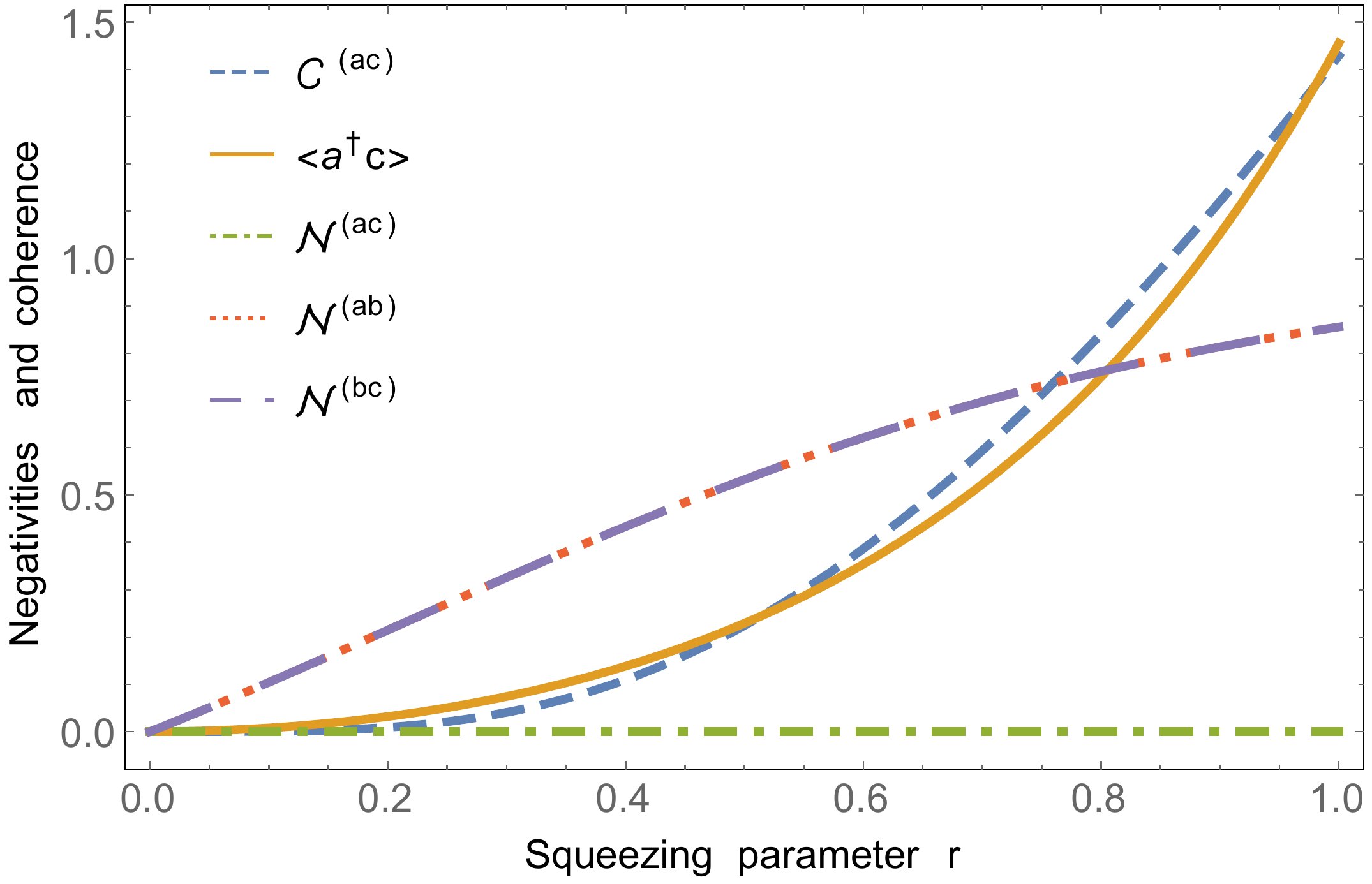}
\caption{
Bipartite negativities of the reduced states $\mathcal{N}^{(ac)}$, $\mathcal{N}^{(ab)}$, $\mathcal{N}^{(bc)}$, coherence measurement $C^{(ac)}$ and interference-based coherence $\langle a^\dagger c\rangle$ of the a-c reduced state vs. squeezing parameter $r=R_{ab}=R_{bc}$ for $\omega_a = 2\pi \times 4.99\operatorname{GHz}$, $\omega_b = 2\pi \times 5 \operatorname{GHz}$, $\omega_c = 2\pi \times 5.01\operatorname{GHz}$ and $T=15\operatorname{mK}$. Note that $\mathcal{N}^{(bc)}=\mathcal{N}^{(ab)}$ due to squeezing in each parametric process separately, while $\mathcal{N}^{(ac)}=0$.}
\label{fig:fig3}
\end{figure}

%------------------------------------------------------------------------------------------------------------------------------------------------------------------------%
We are now ready to discuss some peculiar aspects of the ``ac'' subsystem. We proceed to show that, although the modes a and c have not been directly squeezed and therefore there is no entanglement between them, we still witness the appearance of nontrivial bipartite coherence correlations $\langle a^{\dag}\,c\rangle \neq 0 $. This term can be obtained in a simple way as $\langle a^{\dag}\,c\rangle=\frac{1}{2}\,\sigma_{31}$. We find
\begin{align}
\langle a^{\dag}\,c\rangle_{\rho}=\frac{\delta }{2}. \label{coherence:correlation}
\end{align}
The mechanism by which these correlations are established reminds of the standard which-way information concepts from interferometry. In standard interferometry (or each time we deal with a linear superposition of states) the absence of information about the path that the photon takes (or, equivalently, the information about which specific wave-function within the superposition that constitutes tho total wave-function of the particle is ``actualized'') results in the formation of an interference pattern. In this case, given a boson occupying mode $b$, we cannot know from which downconversion process (corresponding to either pump $\omega_{ab}$ or $\omega_{bc}$) it originates. The first-order coherence  $g^{(1)}_{ac}$ of modes $a$, $c$ can be readily obtained,
\begin{equation}
g^{(1)}_{ac} = \frac{\langle a^{\dag}\,c\rangle}{\sqrt{\langle a^{\dag}\,a\rangle\langle c^{\dag}\,c\rangle}},
\end{equation}
and using $\langle a^{\dag}\,a\rangle = (\alpha -1)/2$ and $\langle c^{\dag}\,c\rangle = (\gamma -1)/2$ as determined from \eqref{final:three:mode:state:structure}
we get
\begin{equation}\label{first:order:coherence}
g^{(1)}_{ac} = \frac{\delta}{\sqrt{(\alpha -1)(\gamma -1)}}.
\end{equation}
Finally,  the single-particle density matrix $\boldsymbol{\rho}^{\bf(1)}_{ac}$ is
\begin{align}
\label{singlemodedensitymatrix1}
\boldsymbol{\rho^{(1)}}_{ac}=
\begin{pmatrix}
\langle a^{\dag}\,a\rangle & \langle a^{\dag}\,c\rangle \\
\langle c^{\dag}\,a\rangle  & \langle c^{\dag}\,c\rangle \\
\end{pmatrix}
= \frac{1}{2} \begin{pmatrix}
\alpha -1 & \delta \\
\delta & \gamma -1 \\
\end{pmatrix} .
\end{align}
From Fig. (\ref{fig:fig3}) we see that a nonzero degree of bipartite coherence between $a$ and $c$ exists, and it increases with the squeezing.
Also the relative entropy of coherence $C^{(ac)}$ in  $\boldsymbol{\sigma}^{(ac)}$ can be calculated from Eqs. (\ref{entropy}-\ref{coherence}), and this quantity is nonvanishing as well.

This underlines the fundamental difference between the two-mode correlations produced by a single pump (which produce entanglement but no bipartite coherence) and those produced between the extremal modes in the double dynamical Casimir effect (which have coherence but no entanglement).

%We conclude by noting that, although in the final state the entanglement between modes $a$ and $c$ is vanishingly small, by measuring the mode $b$ in a homodyne scheme one can induce a finite degree of entanglement between modes $a$ and $c$. This will be demonstrated in the next section.

%------------------------------------------------------------------------------------------------------------------------------------------------------------------------%
\section{Homodyne measurement of the common idler and coherence-to-entanglement conversion}\label{section:meas}
%------------------------------------------------------------------------------------------------------------------------------------------------------------------------%
In this section we compute the resulting state of modes $a$ and $c$ after a \textit{perfect} homodyne detection of mode $b$. In particular, we analyse the coherence and entanglement of the resulting state. In order to reach this goal, we  employ the formalism of homodyne detection developed in Ref. [\onlinecite{spedalieri}]. The technical details of this can be found in Appendix \ref{appendix:one:one} and we omit them here in order to focus on correlations between modes $a$ and $c$ after homodyne detection.

After some lengthy algebra (see also \cite{Cirac:2002}), one has the final state $\boldsymbol{\sigma}^{(ac)}_{out|q}$ of modes $a$ and $c$ after homodyne detection of the quadrature $q$ of mode $b$, which reads
\begin{align}
\boldsymbol{\sigma}^{(ac)}_{{\rm out}|q}=
\begin{pmatrix}
\alpha-\frac{\epsilon^2}{2\beta} & \delta-\frac{\epsilon\zeta}{2\beta} & -\frac{\epsilon^2}{2\beta} & -\frac{\epsilon\zeta}{2\beta}\\
\delta-\frac{\epsilon\zeta}{2\beta} & \gamma-\frac{\zeta^2}{2\beta} & -\frac{\epsilon\zeta}{2\beta} & -\frac{\zeta^2}{2\beta} \\
-\frac{\epsilon^2}{2\beta}&-\frac{\epsilon\zeta}{2\beta} & \alpha-\frac{\epsilon^2}{2\beta} & \delta-\frac{\epsilon\zeta}{2\beta} \\
 -\frac{\epsilon\zeta}{2\beta}& -\frac{\zeta^2}{2\beta}& \delta-\frac{\epsilon\zeta}{2\beta} & \gamma-\frac{\zeta^2}{2\beta}
\end{pmatrix}. \label{eq:outqus}
\end{align}
A simple inspection of the state \eqref{eq:outqus} after homodyne detection allows the identification of the differences with respect to the reduced state $\boldsymbol{\sigma}^{(ac)}$. We see that the coherence $\delta$ of the latter is now decreased. In particular, we expect entanglement to be present in the new state \eqref{eq:outqus} since it contains nonvanishing elements in the upper-right part of the state. To prove analytically that there is entanglement requires lengthy formulas, but we will see later that in the case of frequencies that are very close to each other, analytical insight can be gained.
The smallest symplectic eigenvalue $\tilde{\nu}_-^{{\rm out}|q}$  of the partial transpose of the state \eqref{eq:outqus} can be computed and has a lengthy expression, which we reproduce in Appendix \ref{appendix:two}.

The computation of the negativity of the state \eqref{eq:outqus} after homodyne detection is straightforward and follows step-by-step what has been done above.

\begin{figure}[h!]
	\includegraphics[width=0.5\textwidth]{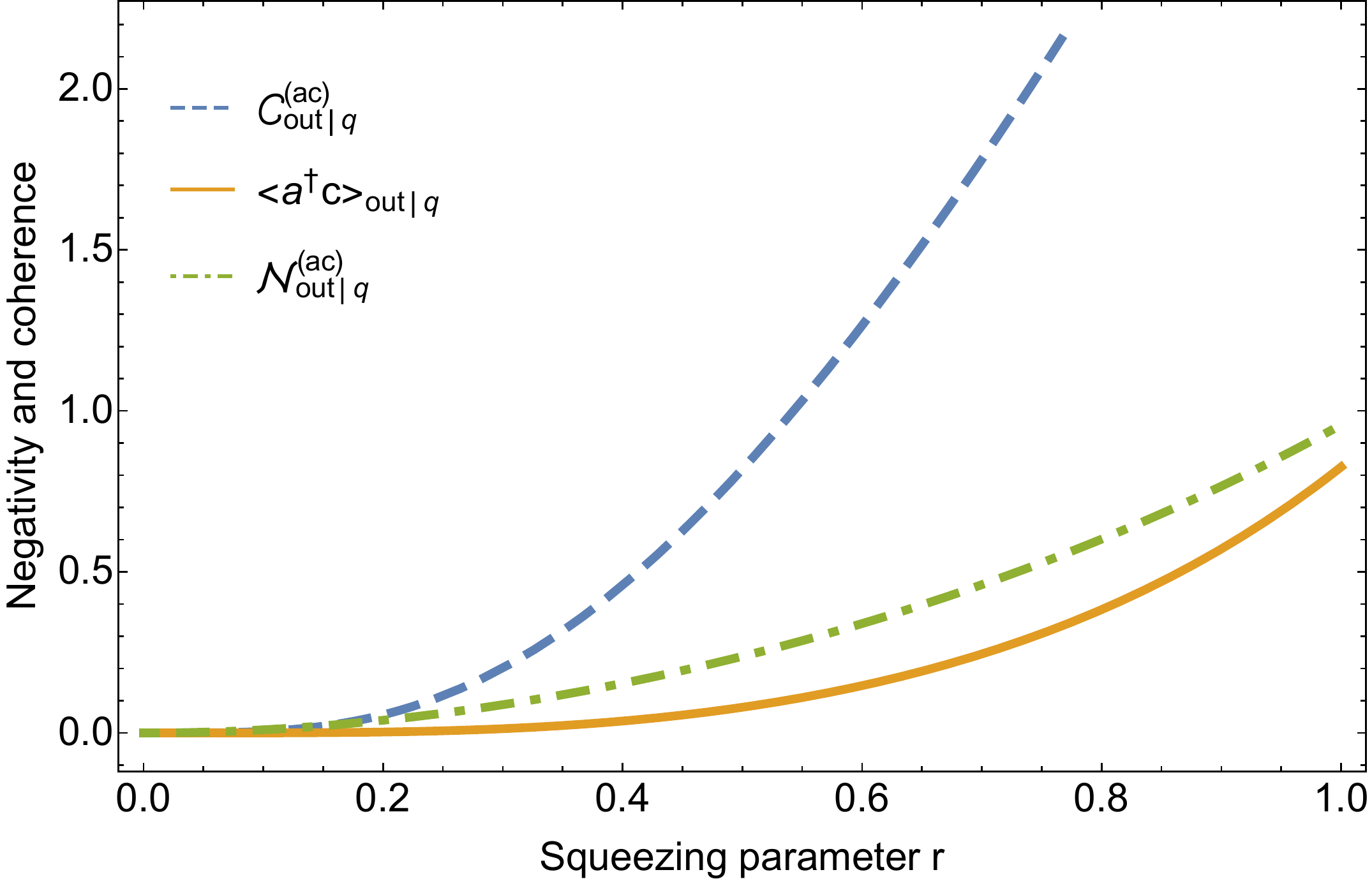}
	\caption{Negativity $\mathcal{ N}^{(ac)}_{{\rm out}|q}$, bipartite coherence $\langle a^\dagger\, c\rangle_{{\rm out}|q}$, and entropy of coherence $ C^{(ac)}_{{\rm out}|q}$  versus squeezing parameter $r=R_{ab}=R_{bc}$ of the state $\boldsymbol{\sigma}^{(ac)}_{{\rm out}|q}$ for $\omega_a = 2\pi \times 4.99\operatorname{GHz}$, $\omega_b = 2\pi \times 5\operatorname{GHz}$, $\omega_c = 2\pi \times 5.01\operatorname{GHz}$ and $T=15\operatorname{mK}$.}
	\label{fig4}
\end{figure}

\begin{figure}[h!]
	\includegraphics[width=0.5\textwidth]{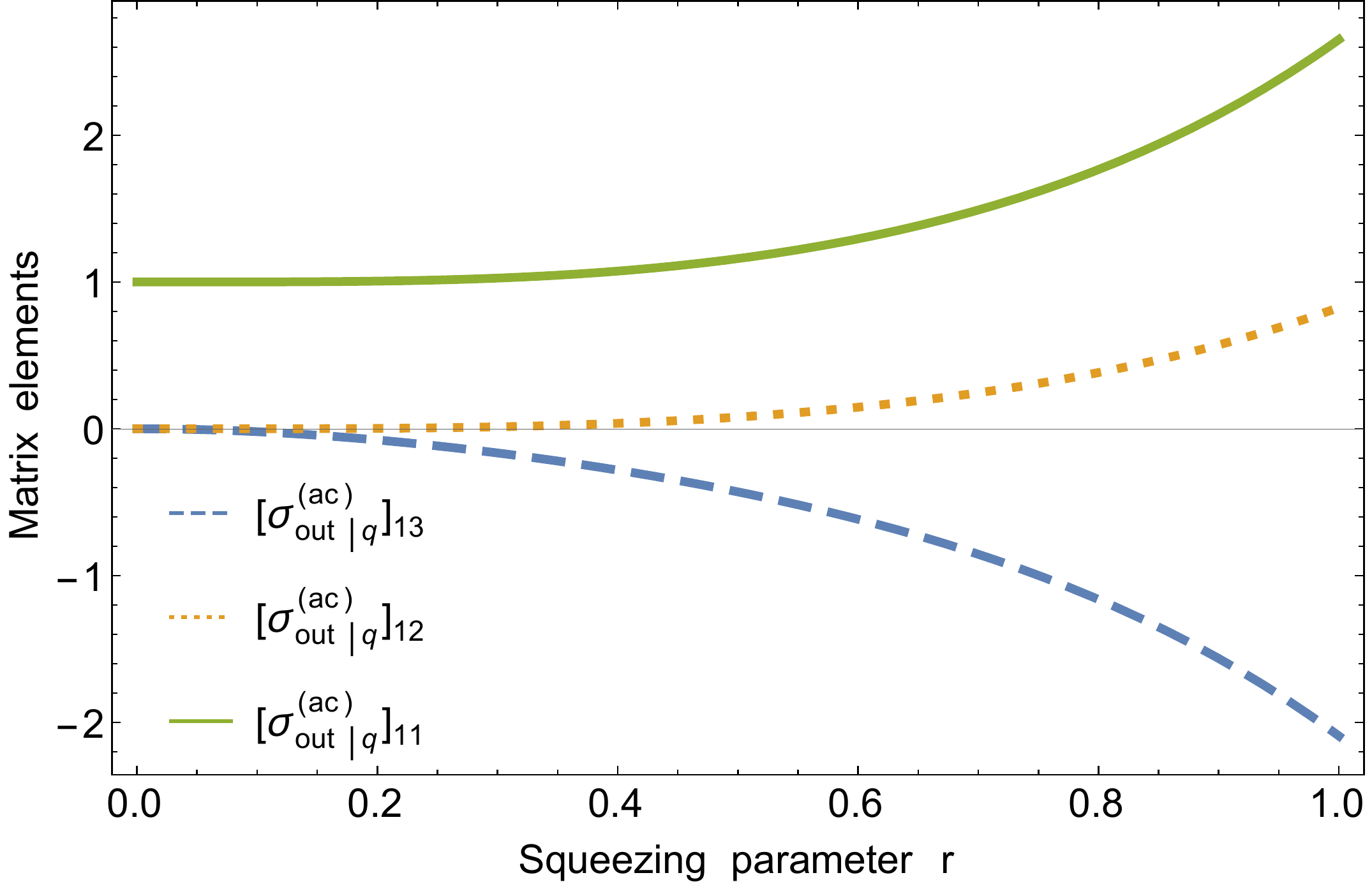}
	\caption{Elements of the state $\boldsymbol{\sigma}^{(ac)}_{{\rm out}|q}$ as a function of the squeezing parameter, for the same parameters as in the previous figures. Here we have $[\boldsymbol{\sigma}^{(ac)}_{{\rm out}|q}]_{11}=[\boldsymbol{\sigma}^{(ac)}_{{\rm out}|q}]_{22}$ and $[\boldsymbol{\sigma}^{(ac)}_{{\rm out}|q}]_{13}$ = $[\boldsymbol{\sigma}^{(ac)}_{{\rm out}|q}]_{14}$=$[\boldsymbol{\sigma}^{(ac)}_{{\rm out}|q}]_{24}$,   $[\boldsymbol{\sigma}^{(ac)}_{{\rm out}|q}]_{12}$.}
	\label{fig5}
\end{figure}

In Figure \ref{fig4} we plot the negativity $\mathcal{N}$, the bipartite coherence $\langle a^\dagger c\rangle$, and the entropy of coherence $C$ as a function of the squeezing parameter $r=R_{ab}=R_{bc}$ for typical values of frequencies and  temperature encountered in experiments with superconducting circuits. The result is that after this projective measurement of quadrature $q$ in the common idler mode $b$, the entanglement becomes nonzero at the expense of a reduction in bipartite coherence. In other words the state has become more squeezed in the two modes but has lost in interferometric visibility. In Figure \ref{fig5} we also compare $\langle a^{\dag}\,c\rangle_{{\rm out}|q}=1/2[\boldsymbol{\sigma}^{(ac)}_{{\rm out}|q}]_{12}$  with the other elements of the correlation matrix corresponding to the state.
Note that the entropy of coherence also changes, but not significantly. This occurs because the entropy of coherence is a measure of the total coherence-like resources available (it quantifies globally how different the state is from a mixed state) but it doesn't contain any information about how these resources are distributed between the states forming the basis (the number of particle eigenstates).
In contrast, the bipartite coherence concentrates on only one resource, the visibility in an interference experiment realized with two modes, and ignores the information about the other possible correlations.

The result can be regarded as a phenomenon of redistribution of tripartite quantum resources into a bipartite resource (bipartite entanglement), when one mode is eliminated by measurement. Similar effects have been studied before with discrete variables in spin chains, where entanglement can be localized between two spins by the measurement of the other spins \cite{Verstraete:2004}. With superconducting circuits, in a tripartite system consisting of two superconducting qubits coupled to a resonator with only one quanta of excitation present, it was shown that a null measurement on the number of particles in the resonator results in the creation of entanglement between the qubits \cite{PhysRevB.78.064503,1742-6596-150-2-022051}, a technique that can be thought of as a particular form of dissipation engineering \cite{PhysRevA.59.2468}.

To get a better grasp of this, let us consider a W state, which is a tripartite state with one excitation distributed in an equal superposition over three modes, $|\mathrm{W}\rangle = 1/\sqrt{3}\left(|001\rangle + |010\rangle + |100\rangle \right)$. These states display a different type of entanglement from GHZ states \cite{GHZ}, and their nonlocal features have been studied intensely \cite{Paraoanu2011}.
It is easy to check that bi-squeezed tripartite states truncated to the subspace of at most two excitations have indeed a W structure \cite{Paraoanu:coherence}. If any of the modes is traced out, the remaining two modes are entangled with concurrence 2/3. Consider now a measurement on any of the three modes: if the result is $0$, then we have projected the W state into a maximally entangled Bell state $(1/\sqrt{2})\left(|01\rangle + |10\rangle \right)$, thus increasing  the concurrence from 2/3 to 1. However, differently from the case of discrete variables, in the case of Gaussian states the resulting covariance matrix does not depend on the value $q$ measured. This feature makes for example impossible to distill Gaussian states with Gaussian operations \cite{PhysRevLett.89.137903} - in sharp contrast with the case of discrete variables where this is possible - since we cannot post-select with respect to the result of the measurement. It is also worth noting that the homodyne detection scheme is optimal for achieving the highest degree of entanglement \cite{Fiurasek:2007,Fiurasek:2008}.

Testing the prediction of entanglement generation between the modes $a$ and $c$ through the measurement of mode $b$ can be done by first setting a $q$-quadrature (in-phase) homodyne detection in the $b$ mode - through the adjustment of the phase of the local oscillator ${\rm (LO)}_b$ - then performing correlation measurements between the $a$-$c$ modes.
Thus, the experimental realization would require only the addition of an analysis channel to the present setups used to measure two-frequency correlations in the dynamical Casimir effect.
%------------------------------------------------------------------------------------------------------------------------------------------------------------------------%
\section{Applications to microwave and optical experimental setups}\label{section:appl}
%------------------------------------------------------------------------------------------------------------------------------------------------------------------------%

The results presented above can be readily applied to existing experimental platforms such as superconducting circuits realized with coplanar waveguide resonators terminated by SQUIDs \cite{Paraoanu:coherence, Wilson:Johansson:2011} and SQUID arrays \cite{Paraoanu:Pnas} as typically designed for operation as microwave parametric amplifiers (see {\it e.g.} \cite{Lahteenmaki2014} for a review) and to optical systems using pumped nonlinear crystals \cite{PhysRevLett.101.130501,PhysRevLett.112.120505}.

We start by noticing that the correlations generated by double parametric pumping depend sensitively on the initial temperature. The other relevant parameter for experiments is the frequency difference between the modes. In order to discuss the competition between temperature and frequency difference in realistic experimental scenarios, we find it convenient for simplifying the notations to introduce the dimensionless frequencies $\Omega_m$ as $\Omega_m:=\frac{\hbar\,\omega_m}{k_B\,T}$, where $m=a,b,c$.

This allows us to rewrite the symplectic eigenvalues $\nu_m$ as $\nu_m=\coth(\Omega_m /2)$, with $m=a,b,c$.
Given this redefinition, we can simplify the notations easily and define the reference dimensionless frequency $\Omega:=\Omega_b$ and introduce the shift in the dimensionless frequencies $\delta\Omega$ through $\Omega-\delta\Omega=\Omega_a$ and $\Omega+\delta\Omega=\Omega_c$. Note that $\delta\Omega/\Omega\leq1$. Let us consider some examples. In superconducting circuits, typical values for frequencies are $\omega = 2\pi \times 5$ GHz and $\delta \omega= 2\pi \times 10$ MHz at $T=15$mK, while for optical systems we have typically $\omega = 2\pi \times 5.64 \times 10^5$ GHz (532 nm) and $\delta \omega = 2 \pi \times 0.95$ GHz at a temperature  $T=300$K.  These numbers translate into $\Omega=15.9$ and $\delta\Omega=0.04$ for superconducting circuits, while $\Omega=89.8$ and $\delta\Omega= 1.5\times10^{-4}$ for optical systems. In superconducting circuits, temperatures higher than 15 mK are also relevant due to imperfect thermalization and nonequilibrium heating effects.

We are therefore interested in the following two scenarios.
\begin{itemize}
	\item[i)] The reference dimensionless frequency $\Omega$ is of the order of unity or smaller and $\delta\Omega/\Omega\ll1$. The three symplectic eigenvalues are $\nu_{\pm}=\coth(\Omega/2)[1\mp\delta\Omega /\sinh(\Omega)]$ and $\nu=\coth(\Omega /2)$ to first order in $\mathcal{O}(\delta\Omega /\Omega)$.
	\item[ii)] The lowest frequency satisfies $\Omega-\delta\Omega\gg 1$ or, in other words, the thermal energy available is extremely low compared to the energy cost of each excitation. This implies that we can safely set $\nu_m=1$ for each mode.
\end{itemize}
We proceed with the analysis of each scenario.

%------------------------------------------------------------------------------------------------------------------------------------------------------------------------%
\subsection{Modes with closely separated frequencies}
%------------------------------------------------------------------------------------------------------------------------------------------------------------------------%
Let us assume that $\Omega$ is of the order of unity or smaller and $\delta\Omega/\Omega\ll1$. This occurs in superconducting circuits when the temperatures are (much) higher than $T=10\operatorname{mK}$. This is the regime where the initial mixedness due to temperature can be important and the frequencies can be regarded as close enough.

Let us start by noting that in superconducting qubits one typically has $\delta\omega/\omega=2\times10^{-3}\ll1$ \cite{Paraoanu:coherence}. Then
\begin{align}
\nu_{\pm}=\coth\left(\frac{\Omega}{2}\right)\left(1\mp \frac{\delta\Omega}{\Omega}\frac{\Omega}{\sinh \Omega}\right),
\end{align}
where the factor $\Omega /\sinh \Omega$, in this regime, is a number close to unity.

In this case, we see that the symplectic eigenvalues $\nu_{\pm}$ and $\nu$ coincide to very good approximation and we ignore contributions of the order $\mathcal{O}(\delta\Omega /\Omega)$. We can therefore obtain the elements of the covariance matrix $\boldsymbol{\sigma}$ of the final state \eqref{final:three:mode:state:structure} which, to lowest order, reduce to
\begin{align}
\alpha=&\nu\,\cosh(2\,r_{ab}),\nonumber\\
\beta=&\nu\,\left[\cosh(2\,r_{ab})\,\cosh^2(r_{bc}) + \sinh^2(r_{bc})\right],\nonumber\\
\gamma=&\nu\,\left[\cosh(2\,r_{ab})\,\sinh^2(r_{bc}) + \cosh^2(r_{bc})\right],\nonumber\\
\delta=&\nu\,\sinh(2\,r_{ab})\,\sinh(r_{bc}),\nonumber\\
\epsilon=&\nu\,\sinh(2\,r_{ab})\,\cosh(r_{bc}),\nonumber\\
\zeta=&\nu\,\cosh^2(r_{ab})\,\sinh(2\,r_{bc}).
\end{align}
These simplified terms allow us to obtain a better analytical understanding of our system, since they dramatically reduce the algebra involved when computing the relevant figures of merit.

In particular, we can focus on the smallest symplectic eigenvalues of the reduced states, since they contain all the necessary information to determine quantum correlations. It is easy to check that the smallest symplectic eigenvalues $\tilde{\nu}_-^{(ab)}$, $\tilde{\nu}_-^{(bc)}$ and $\tilde{\nu}_-^{(ac)}$ of the reduced states of modes (ab), (bc) and (ac) respectively are
\begin{widetext}
\begin{align}\label{smallest:symplectic:eigenvalues:same:frequency}
\tilde{\nu}_-^{(ab)}=&\nu\,\left[1+2\,{\rm sh}^2_{ab}+{\rm sh}^2_{bc}+{\rm sh}^2_{ab}\,{\rm sh}^2_{bc}-\sqrt{4\,{\rm sh}^2_{ab}+4\,{\rm sh}^4_{ab}+{\rm sh}^4_{bc}+4\,{\rm sh}^2_{ab}\,{\rm sh}^2_{bc}+2\,{\rm sh}^2_{ab}\,{\rm sh}^4_{bc}+4\,{\rm sh}^4_{ab}\,{\rm sh}^2_{bc}+{\rm sh}^4_{ab}\,{\rm sh}^4_{bc}}\right],\nonumber\\
\tilde{\nu}_-^{(bc)}=&\nu\,\left[1+{\rm sh}^2_{ab}+2\,{\rm sh}^2_{bc}+2\,{\rm sh}^2_{ab}\,{\rm sh}^2_{bc}-\sqrt{{\rm sh}^4_{ab}+4(1+{\rm sh}^2_{ab})^2\,{\rm sh}^2_{bc}\,(1+{\rm sh}^2_{bc})}\right],\nonumber\\
\tilde{\nu}_-^{(ac)}=&\nu\,\left[\sqrt{1+2\,{\rm sh}^2_{ab}+2\,{\rm sh}^2_{bc}+{\rm sh}^4_{ab}+{\rm sh}^4_{bc}+2\,{\rm sh}^2_{ab}\,{\rm sh}^4_{bc}+{\rm sh}^4_{ab}\,{\rm sh}^4_{bc}}-|{\rm sh}^2_{ab}-{\rm sh}^2_{bc}-{\rm sh}^2_{ab}\,{\rm sh}^2_{bc}|\right].
\end{align}
\end{widetext}
These smallest symplectic eigenvalues are not always smaller than one, {\it i.e.} it is not always guaranteed that there is entanglement between the modes in the reduced subsystem. We see that the conditions $\tilde{\nu}_-^{(nm)}<1$ for the existence of the entanglement in the reduced states (ab),(bc) and (ac) are, respectively,
\begin{align}\label{conditions:same:frequency:satisfied:bla}
{\rm sh}^2_{ab}>&\left(\frac{\nu-1}{2\,\nu}\right)^2+\frac{\nu-1}{2\,\nu}\,\left[2\,{\rm sh}^2_{ab}+{\rm sh}^2_{bc}+{\rm sh}^2_{ab}\,{\rm sh}^2_{bc}\right],\nonumber\\
{\rm sh}^2_{bc}>&\left(\frac{\nu-1}{2\,\nu}\right)^2+\frac{\nu-1}{2\,\nu}\,\left[{\rm sh}^2_{ab}+2\,{\rm sh}^2_{bc}+2\,{\rm sh}^2_{ab}\,{\rm sh}^2_{bc}\right],
\end{align}
and
\begin{align}\label{conditions:same:frequency:not:satisfied:bla}
\frac{1-\nu^2}{\nu^2}>&\frac{2}{\nu}\,|{\rm sh}^2_{ab}-{\rm sh}^2_{bc}-{\rm sh}^2_{ab}\,{\rm sh}^2_{bc}|\nonumber\\
&+2\,{\rm sh}^2_{ab}+2\,{\rm sh}^2_{bc}+2\,{\rm sh}^2_{ab}\,{\rm sh}^2_{bc}+2\,{\rm sh}^4_{ab}\,{\rm sh}^2_{bc}.
\end{align}
We notice that the two conditions in \eqref{conditions:same:frequency:satisfied:bla} are not always satisfied, which means that there is need for a finite amount of squeezing before any correlation can be established. This is in agreement with previous work that has analysed the interplay of initial mixedness, due to temperature, and squeezing \cite{Bruschi:Friis:2013}. The exact value of the squeezings, as a function of the initial mixedness $\nu$, at which entanglement is created can be found  by looking at the point of saturation of the inequality in these two conditions.

We finally note that the last condition \eqref{conditions:same:frequency:not:satisfied:bla} is never satisfied, since $\nu\geq1$ and the right hand side is always positive. This is expected from the form of the final state  of modes $a$ and $c$. This means that there is never entanglement between these two modes.

We can also look at the final state (ac) \textit{after} homodyne detection in this regime. As anticipated before, we are now able to show that the state (\ref{eq:outqus}) is entangled. We start by noting that the two reduced states $\boldsymbol{\sigma}^{(a)}_{{\rm out}|q}$ and $\boldsymbol{\sigma}^{(c)}_{{\rm out}|q}$
 of modes $a$ and $c$, and the $2\times2$ correlation block $\boldsymbol{\sigma}^{(\text{corr})}_{{\rm out}|q}$, read
\begin{align}\label{reduced:states:after:homodyne}
\boldsymbol{\sigma}^{(a)}_{{\rm out}|q}=
\begin{pmatrix}
\alpha-\frac{\epsilon^2}{2\beta} & -\frac{\epsilon^2}{2\beta} \\
-\frac{\epsilon^2}{2\beta} & \alpha-\frac{\epsilon^2}{2\beta}
\end{pmatrix},\nonumber\\
\boldsymbol{\sigma}^{(c)}_{{\rm out}|q}=
\begin{pmatrix}
 \gamma-\frac{\zeta^2}{2\beta} & -\frac{\zeta^2}{2\beta} \\
 -\frac{\zeta^2}{2\beta}& \gamma-\frac{\zeta^2}{2\beta}
\end{pmatrix},\nonumber\\
\boldsymbol{\sigma}^{(\text{corr})}_{{\rm out}|q}=
\begin{pmatrix}
 \delta-\frac{\epsilon\zeta}{2\beta} &  -\frac{\epsilon\zeta}{2\beta}\\
-\frac{\epsilon\zeta}{2\beta}  & \delta-\frac{\epsilon\zeta}{2\beta}
\end{pmatrix}.
\end{align}
We then introduce the \textit{local symplectic invariants} $a^2,b^2$ and $c_+\,c_-$, defined as $a^2:=\text{det}(\boldsymbol{\sigma}^{(a)}_{{\rm out}|q})$, $b^2:=\text{det}(\boldsymbol{\sigma}^{(b)}_{{\rm out}|q})$ and $c_+\,c_-:=\text{det}(\boldsymbol{\sigma}^{(\text{corr})}_{{\rm out}|q})$. We choose to keep the notation for the local invariants as in standard reference. The local invariants $a$ and $b$ here are not to be confused with the mode operators. We notice that a two-mode entangled state is \textit{symmetric} if, in a decomposition of this form, $a^2=b^2$ (see \cite{Adesso:Ragy:2014}). In our case we have
\begin{align}
a^2=b^2=&\nu^2\,\frac{(1+2\,{\rm sh}^2_{bc}+2\,{\rm sh}^2_{ab}\,{\rm sh}^2_{bc})\,(1+2\,{\rm sh}^2_{ab})}{1+2\,{\rm sh}^2_{ab}+2\,{\rm sh}^2_{bc}+2\,{\rm sh}^2_{ab}\,{\rm sh}^2_{bc}}, \nonumber\\
c_+\,c_-=&-4\,\nu^2\,\frac{(1+{\rm sh}^2_{ab})\,{\rm sh}^2_{ab}\,{\rm sh}^2_{bc}}{1+2\,{\rm sh}^2_{ab}+2\,{\rm sh}^2_{bc}+2\,{\rm sh}^2_{ab}\,{\rm sh}^2_{bc}}.
\end{align}
which confirms that we have a symmetric two-mode Gaussian state.
It is known that every two mode symmetric Gaussian state is equivalent to a two-mode squeezed state up to local operations \cite{Adesso:Ragy:2014}. This implies that we can anticipate squeezing between modes $a$ and $c$, which we proceed to compute.

Now we can compute the smallest symplectic eigenvalue $\tilde{\nu}_-^{{\rm out}|q}$ of the partial transpose in order to quantify the squeezing between the two modes. This can be done by employing a known relation between the local symplectic eigenvalues, which has the expression $2\,(\tilde{\nu}_-^{{\rm out}|q})^2=\tilde{\Delta}-\sqrt{\tilde{\Delta}^2-4\,\text{det}(\boldsymbol{\sigma}^{(ac)}_{{\rm out}|q})}$, where we have introduced $\tilde{\Delta}:=a^2+b^2-2\,c_+\,c_-$ for convenience of presentation.
In our case this expression simplifies to
\begin{align}
(\tilde{\nu}_-^{{\rm out}|q})^2=a^2-c_+\,c_--\sqrt{(a^2-c_+\,c_-)^2-\text{det}(\boldsymbol{\sigma}^{(ac)}_{{\rm out}|q})},
\end{align}
which allows us immediately find the condition, analogous to \eqref{conditions:same:frequency:satisfied:bla} and  \eqref{conditions:same:frequency:not:satisfied:bla}, for the existence of the entanglement in this case. We have
\begin{align}\label{conditions:same:frequency:homodyne}
\frac{(1+{\rm sh}^2_{ab})\,{\rm sh}^2_{ab}\,{\rm sh}^2_{bc}}{1+2\,{\rm sh}^2_{ab}+2\,{\rm sh}^2_{bc}+2\,{\rm sh}^2_{ab}\,{\rm sh}^2_{bc}}>&\left(\frac{\nu^2-1}{4\,\nu}\right)^2.
\end{align}
We can immediately see that, if $r_{ab}=0$ or $r_{bc}=0$, then the condition \eqref{conditions:same:frequency:homodyne} is never satisfied and there is no entanglement in the state $(ac)$ after homodyne detection, as expected.

In Appendix \ref{appendix:two} we provide an explicit expression for the smallest symplectic eigenvalue $\tilde{\nu}_-^{{\rm out}|q}$ of the partial transpose.

For completeness, we can obtain explicit formulas for the behavior of the first order coherence $g^{(1)}_{ac} $ in this regime. This can be done using Eq. \eqref{first:order:coherence} and the expressions above. We find
\begin{align}
g^{(1)}_{ac}=2\,\nu\,\frac{{\rm sh}_{ab}\,{\rm sh}_{bc}\,{\rm ch}_{ab}}{\sqrt{(\nu-1+2\,\nu\,{\rm sh}^2_{ab})(\nu-1+2\,\nu\,{\rm sh}^2_{bc}\,{\rm ch}^2_{bc})}}.
\end{align}

%------------------------------------------------------------------------------------------------------------------------------------------------------------------------%
\subsection{Low temperatures}
%------------------------------------------------------------------------------------------------------------------------------------------------------------------------%
We can now investigate the ``low enough'' temperature regime. We have seen that $\Omega\gg1$, both in superconducting circuits and in optical cavities. In particular, $\Omega=15$ for microwaves and $\Omega=84.6$ for optical cavities. This implies that, in both scenarios, $\coth(\Omega_m)\sim1-2\exp[-\Omega_m]$ which, for all purposes, is unity. The discussion in this section applies as long as $\delta\Omega/\Omega\ll1$ as well.

The consequence of these considerations is that we can safely set $\nu=1$ in the results of the previous section. This implies that the symplectic eigenvalues now read, to leading order,
\begin{widetext}
\begin{align}
\tilde{\nu}_-^{(ab)}=&1+2\,{\rm sh}^2_{ab}+{\rm sh}^2_{bc}+{\rm sh}^2_{ab}\,{\rm sh}^2_{bc}-\sqrt{4\,{\rm sh}^2_{ab}+4\,{\rm sh}^4_{ab}+{\rm sh}^4_{bc}+4\,{\rm sh}^2_{ab}\,{\rm sh}^2_{bc}+2\,{\rm sh}^2_{ab}\,{\rm sh}^4_{bc}+4\,{\rm sh}^4_{ab}\,{\rm sh}^2_{bc}+{\rm sh}^4_{ab}\,{\rm sh}^4_{bc}},\nonumber\\
\tilde{\nu}_-^{(bc)}=&1+{\rm sh}^2_{ab}+2\,{\rm sh}^2_{bc}+2\,{\rm sh}^2_{ab}\,{\rm sh}^2_{bc}-\sqrt{{\rm sh}^4_{ab}+4(1+{\rm sh}^2_{ab})^2\,{\rm sh}^2_{bc}\,(1+{\rm sh}^2_{bc})},\nonumber\\
\tilde{\nu}_-^{(ac)}=&\sqrt{1+2\,{\rm sh}^2_{ab}+2\,{\rm sh}^2_{bc}+{\rm sh}^4_{ab}+{\rm sh}^4_{bc}+2\,{\rm sh}^2_{ab}\,{\rm sh}^4_{bc}+{\rm sh}^4_{ab}\,{\rm sh}^4_{bc}}-|{\rm sh}^2_{ab}-{\rm sh}^2_{bc}-{\rm sh}^2_{ab}\,{\rm sh}^2_{bc}|,
\end{align}
\end{widetext}
which also implies that the conditions \eqref{conditions:same:frequency:satisfied:bla}, \eqref{conditions:same:frequency:not:satisfied:bla} and \eqref{conditions:same:frequency:homodyne} for the existence of entanglement reduce to
\begin{align}\label{conditions:zero:temperature}
{\rm sh}^2_{ab}>&0,\nonumber\\
1>&\frac{1}{2}\,(1-{\rm th}^2_{ab})\,\frac{1-{\rm th}^2_{bc}}{1+{\rm th}^2_{bc}},\nonumber\\
0>&\frac{2}{\nu}\,|{\rm sh}^2_{ab}-{\rm sh}^2_{bc}-{\rm sh}^2_{ab}\,{\rm sh}^2_{bc}|\nonumber\\
&+2\,{\rm sh}^2_{ab}+2\,{\rm sh}^2_{bc}+2\,{\rm sh}^2_{ab}\,{\rm sh}^2_{bc}+2\,{\rm sh}^4_{ab}\,{\rm sh}^2_{bc}\nonumber\\
{\rm sh}^2_{ab}\,{\rm sh}^2_{bc}>&0.
\end{align}
The first two and the last conditions in \eqref{conditions:zero:temperature} are always satisfied. This can be easily explained by the fact that there is no initial mixedness that competes with the establishment of correlations between the different modes. In Figures \ref{fig:fig2} we see that that these conditions are always satisfied, which is equivalent to the fact that the curve for this case is always positive, except at the origin.
For completeness, we can use \eqref{number:expectation:value:equations} to find the average excitation in this low temperature regime. We find
\begin{align}\label{number:expectation:value:equations:vacuum:state}
\langle a^{\dag}\,a\rangle=&{\rm sh}^2_{ab}\nonumber\\
\langle b^{\dag}\,b\rangle=&{\rm sh}^2_{ab}+{\rm sh}^2_{bc}+{\rm sh}^2_{ab}\,{\rm sh}^2_{bc}\nonumber\\
\langle c^{\dag}\,c\rangle=&{\rm sh}^2_{bc}\,\left(1+{\rm sh}^2_{ab}\right).
\end{align}
The third condition in \eqref{conditions:zero:temperature} is never satisfied, again, as expected. This implies that in the (ac) state after homodyne there are genuine correlations irrespectively of the amount of initial squeezing. This is surprising, since one can argue that, after the application of the two mode squeezing operators on modes $a$ and $b$, the reduced state of $a$ is a thermal state with local temperature $T_b$ determined by
\begin{align}
T_b=\frac{\hbar\,\omega}{k_B}\frac{1}{\ln\left(\frac{1+{\rm sh}^2_{ab}+{\rm sh}^2_{bc}+{\rm sh}^2_{ab}\,{\rm sh}^2_{bc}}{{\rm sh}^2_{ab}+{\rm sh}^2_{bc}+{\rm sh}^2_{ab}\,{\rm sh}^2_{bc}}\right)},
\end{align}
which can be derived by equating $\nu_b=\coth(\frac{\Omega_b}{2})=2\,\langle b^{\dag}\,b\rangle+1$, see (\ref{number:expectation:value:equations:vacuum:state}). This concludes our analysis of the low temperature regime.

We can calculate in the same way the behavior of the first order coherence $g^{(1)}_{ac} $ in this regime, employing Eq. \eqref{first:order:coherence} and the expressions derived so far. In this regime we have to be careful of how the limits are taken. As long as $r_{ab}$ and $r_{bc}$ are (possibly small) but finite, we find
\begin{align}
g^{(1)}_{ac}=1+\frac{1}{2}\,\left(\frac{1}{{\rm sh}^2_{ab}}+\frac{1}{{\rm sh}^2_{bc}}\frac{1}{{\rm ch}^2_{ab}}\right)\,e^{-\Omega}.
\end{align}
Note that, due to issues that arise in multi parameter perturbation theory, the above formula \textit{cannot} be applied when $r_{ab}$ or $r_{bc}$ are smaller than $e^{-\Omega}$. Instead, a case-by-case study must be performed in order to establish which parameters are perturbative, and which are not \cite{Safranek:Kholrus:2015}.

We emphasize that the induced correlations discussed here can be regarded as an information-processing resource; in the case of entanglement this has been known in quantum information science for a long time, while in the case of coherence it has been only recently started to be appreciated \cite{Streltsov:Adesso:2016}, opening the way to applications in quantum technologies. In this subsection we have evaluated these quantities under experimentally relevant conditions corresponding to initial mixedness due to finite temperature and with closely separated mode frequencies.

%------------------------------------------------------------------------------------------------------------------------------------------------------------------------%
\section{Conclusions}
%---------------------------------------------------------------------------------------------------------------------z---------------------------------------------------%
We have studied a specific tripartite state of interest for physical implementations in the laboratory, namely a bi-squeezed state. This state can be obtained by applying simultaneous two-mode squeezing between two pairs of modes which share a common third one. We have employed techniques from continuous variables to compute analytically most quantities of interest, such as genuine tri-partite and bipartite entanglement, as well as the purity of all subsystems. We have also analysed the effect of homodyne detection of the common mode, which can be of importance within the development of future technologies. We have found that the modes acquire squeezing-type quantum correlations (nonzero entanglement) after the homodyne detection at the expense of a reduction in bipartite coherence, a phenomenon that can be seen as a redistribution of quantum resources. Finally. we analyzed scenarios of relevance for concrete applications, such as low temperatures or modes with very close frequencies. These situations occur in experiments with superconducting circuits as well as in optical setups aimed at developing the next generation of quantum technologies based on continuous variables.

%------------------------------------------------------------------------------------------------------------------------------------------------------------------------%
\section*{Acknowledgments}
%------------------------------------------------------------------------------------------------------------------------------------------------------------------------%
We thank Marcus Huber, Pertti Hakonen, and Gerardo Adesso for useful comments and discussions. D.E.B. acknowledges hospitality from the University of Vienna and the Hebrew University of Jerusalem, where part of this project was done. Financial support from Fundaci\'on General CSIC (Programa ComFuturo) is acknowledged by C.S. G.S.P thanks FQXi, Centre of Quantum Engineering at Aalto University (project QMET), and the Academy of Finland (project 263457 and project 25020 - Centre of Excellence ``Low Temperature Quantum Phenomena and Devices) for financial support.

%------------------------------------------------------------------------------------------------------------------------------------------------------------------------%
\appendix
%------------------------------------------------------------------------------------------------------------------------------------------------------------------------%
%%
%------------------------------------------------------------------------------------------------------------------------------------------------------------------------%
\section{Bi-squeezed tripartite Gaussian states}\label{appendix:zero}
%------------------------------------------------------------------------------------------------------------------------------------------------------------------------%
We start by analysing the subset $\{G_{ab},G_{bc},B_{ac}\}$ of all the possible $21$ Hermitian operators that are quadratic in the creation and annihilation operators of the modes $a,b,c$ (or, equivalently, in the quadrature oprators). Here we have defined
\begin{align}
G_{ab}:=&a^{\dag}\,b^{\dag}+a\,b,\nonumber\\
G_{bc}:=&b^{\dag}\,c^{\dag}+b\,c,\nonumber\\
B_{ac}:=&i\,\left[a\,c^{\dag}-c\,a^{\dag}\right].
\end{align}
It is easy to check that the operators $G_{ab},G_{bc}$ and $B_{ac}$ form a closed sub-Lie algebra of the full algebra. In fact $[G_{ab},G_{bc}]=-i\,B_{ac}$, $[B_{ac},G_{ab}]=i\,G_{bc}$ and $[B_{ac},G_{bc}]=-i\,G_{ab}$.

Squeezing modes $a$ and $b$ at the same time as modes $b$ and $c$, with parameters $R_{ab}$ and $R_{bc}$ respectively can be done through the unitary operator
\begin{align}\label{sorin:operator}
U=e^{i\,\left[R_{ab}\,G_{ab}+R_{bc}\,G_{bc}\right]}.
\end{align}
It has been shown, see \cite{Bruschi:Lee:2013}, that the operator \eqref{sorin:operator} can be written as
\begin{align}
U=e^{i\,\theta_{ac}\,B_{ac}}\,e^{i\,r_{ab}\,G_{ab}}\,e^{i\,r_{bc}\,G_{bc}},
\end{align}
where the real functions $r_{ab},r_{bc}$ and $\theta_{ac}$ depend on $R_{ab}$ and $R_{bc}$.

It is possible to find $r_{ab},r_{bc}$ and $\theta_{ac}$ as functions of $R_{ab}$ and $R_{bc}$. To do this we introduce
\begin{align}\label{sorin:operator:x}
U(x):=e^{i\,\left[R_{ab}\,G_{ab}+R_{bc}\,G_{bc}\right]\,x},
\end{align}
where we notice that $U(1)$ just coincides with the operator \eqref{sorin:operator} we are interested in. We use the techniques introduced in \cite{Bruschi:Lee:2013}, which prescribe to perform differentiation with respect to $x$ on the left and right side of \eqref{sorin:operator:x} and then multiply both sides on the right by $U^{\dag}(x)$. We obtain the main differential equation
\begin{widetext}
\begin{align}\label{differential:equation}
R_{ab}\,G_{ab}+R_{bc}\,G_{bc}=\dot{\theta}_{ac}\,B_{ac}+\dot{r}_{ab}\,e^{i\,\theta_{ac}\,B_{ac}}\,G_{ab}\,e^{-i\,\theta_{ac}\,B_{ac}}+\dot{r}_{bc}\,e^{i\,\theta_{ac}\,B_{ac}}\,e^{i\,r_{ab}\,G_{ab}}\,G_{bc}\,e^{-i\,r_{ab}\,G_{ab}}\,e^{-i\,\theta_{ac}\,B_{ac}},
\end{align}
\end{widetext}
which provide us with the functions $r_{ab}(x),r_{bc}(x)$ and $\theta_{ac}(x)$, as a function of $R_{ab},R_{bc}$ and $x$. Notice that the dot stands for derivative with respect to $x$. Finally, we need to set $x=1$ in order to find the parameters $r_{ab},r_{bc}$ and $\theta_{ac}$ that we are looking for.

Using the fact that
\begin{align}
e^{i\,r_{ab}\,G_{ab}}\,G_{bc}\,e^{-i\,r_{ab}\,G_{ab}}=&\cosh r_{ab}\,G_{bc}+\sinh r_{ab}\,B_{ac},\nonumber\\
e^{i\,r_{bc}\,G_{bc}}\,B_{ac}\,e^{-i\,r_{bc}\,G_{bc}}=&\cosh r_{bc}\,B_{ac}-\sinh r_{bc}\,G_{ab},\nonumber\\
e^{i\,r_{ab}\,G_{ab}}\,B_{ac}\,e^{-i\,r_{ab}\,G_{ab}}=&\cosh r_{ab}\,B_{ac}+\sinh r_{ab}\,G_{bc},\nonumber\\
e^{i\,\theta_{ac}\,B_{ac}}\,G_{ab}\,e^{-i\,r_{ac}\,B_{ac}}=&\cos \theta_{ac}\,G_{ab}-\sin \theta_{ac}\,G_{bc},\nonumber\\
e^{i\,\theta_{ac}\,B_{ac}}\,G_{bc}\,e^{-i\,r_{ac}\,B_{ac}}=&\cos \theta_{ac}\,G_{bc}+\sin \theta_{ac}\,G_{ab},
\end{align}
we obtain the main differential equations
\begin{align}\label{main:differential:equations}
\dot{r}_{ab}\,\cos\theta_{ac}+\dot{r}_{bc}\,\sin\theta_{ac}\,\cosh r_{ab}=&R_{ab},\nonumber\\
-\dot{r}_{ab}\,\sin\theta_{ac}+\dot{r}_{bc}\,\cos\theta_{ac}\,\cosh r_{ab}=&R_{bc},\nonumber\\
\dot{\theta}_{ac}+\dot{r}_{bc}\,\sinh r_{ab}=&0.
\end{align}
Let us introduce
\begin{align}
\rho:=&\sqrt{R_{ab}^2+R_{bc}^2},\nonumber\\
\tan \phi:=&\frac{R_{bc}}{R_{ab}}.
\end{align}
We can now rewrite the main differential equations \eqref{main:differential:equations} as
\begin{align}\label{main:differential:equations:better:form}
\dot{r}_{ab}=&\rho\,\cos(\phi+\theta_{ac}),\nonumber\\
\dot{r}_{bc}\,\cosh r_{ab}=&\rho\,\sin(\phi+\theta_{ac}),\nonumber\\
\dot{\theta}_{ac}+\dot{r}_{bc}\,\sinh r_{ab}=&0.
\end{align}
Combining the equations in \eqref{main:differential:equations:better:form} we obtain
\begin{align}
\dot{\theta}_{ac}\cot(\phi+\theta_{ac})=-\dot{r}_{ab}\,\tanh r_{ab},
\end{align}
which can be written as $\frac{d}{d\,x}\ln\sin(\phi+\theta_{ac})=-\frac{d}{d\,x}\ln\cosh r_{ab}$.
This gives the following important relation
\begin{align}
\sin(\phi+\theta_{ac})=\frac{\sin\phi}{\cosh r_{ab}},
\end{align}
where we have used the initial conditions $r_{ab}(x=0)=0$ and $\theta_{ac}(x=0)=0$. Notice also that $\phi$ is defined in terms of $R_{ab}$ and $R_{bc}$ and does not depend on $x$.

Using this equation we obtain the first important relation
\begin{align}\label{first:relation}
\sinh r_{ab}=\cos\phi\,\sinh(\rho\,x),
\end{align}
which immediately allows us to find
\begin{align}\label{second:relation}
\sin(\phi+\theta_{ac})=\frac{\sin\phi}{\sqrt{1+\cos^2\phi\,\sinh^2(\rho\,x)}}.
\end{align}
Finally, combining all equations we obtain $\dot{r}_{bc}=\rho\frac{\sin\phi}{1+\cos^2\phi\,\sinh^2(\rho\,x)}$,
which leads to
\begin{align}\label{third:relation}
r_{bc}=\frac{1}{2}\ln\left[\frac{1+\sin\phi\,\tanh(\rho\,x)}{1-\sin\phi\,\tanh(\rho\,x)}\right].
\end{align}
We are finally in the position to obtain the desired relations between $r_{ab},r_{bc}$ and $\theta_{ac}$. All we need to do is set $x=1$ in the main relations \eqref{first:relation},\eqref{second:relation} and \eqref{third:relation}, and invert them. We find
\begin{align}\label{final:result}
r_{ab}=&\ln\left(\cos\phi\,\sinh\rho+\sqrt{1+\cos^2\phi\,\sinh^2\rho}\right),\nonumber\\
r_{bc}=&\frac{1}{2}\ln\left(\frac{1+\sin\phi\,\tanh\rho}{1-\sin\phi\,\tanh\rho}\right),\nonumber\\
\theta_{ac}=&\arctan\left(\frac{\tan\phi}{\cosh\rho}\right)-\phi.
\end{align}
Next, we can make the following checks. Let $R_{bc}=0$, which implies $\rho=R_{ab}$ and $\phi=0$. Then $r_{ab}=R_{ab}$, $r_{bc}=0$ and $\theta_{ac}=0$ as expected.
Now let $R_{ab}=0$, which implies $\rho=R_{bc}$ and $\phi=\pi/2$. Then $r_{ab}=0$, $r_{bc}=R_{bc}$ and $\theta_{ac}=0$, again, as expected.

%------------------------------------------------------------------------------------------------------------------------------------------------------------------------%
\section{Elements of the covariance matrix of the final state}\label{appendix:one}
%------------------------------------------------------------------------------------------------------------------------------------------------------------------------%
Here we reproduce the entries of the final state \eqref{final:three:mode:state:structure}, given that the initial state is thermal, with Williamson form $\boldsymbol{\sigma}_{th}=\text{diag}(\nu_a,\nu_b,\nu_c,\nu_a,\nu_b,\nu_c)$. The algebra necessary to obtain them is straightforward, although cumbersome and not-illuminating. For this reason we present the final results only. We have
\begin{widetext}
\begin{align}
\alpha:=&(\nu_a + (\nu_c-\nu_a)\,\sin^2\,\theta_{ac})\,\cosh^2 r_{ab} + \nu_b\,\sinh^2\,r_{ab},\nonumber\\
\beta:=&\nu_b\,\cosh^2r_{ab}\,\cosh^2r_{bc}+ (\nu_a + (\nu_c-\nu_a)\,\sin^2\,\theta_{ac})\,\sinh^2r_{ab}\,\cosh^2r_{bc} - \frac{(\nu_c - \nu_a)}{2}\,\sin(2\,\theta_{ac})\,\sinh r_{ab}\,\sinh (2\,r_{bc}) \nonumber\\
&+ (\nu_c-(\nu_c-\nu_a)\,\sin^2\,\theta_{ac})\,\sinh^2r_{bc},\nonumber\\
\gamma:=&(\nu_c -(\nu_c- \nu_a)\,\sin^2\,\theta_{ac})\,\cosh^2r_{bc}-\frac{(\nu_c - \nu_a)}{2}\,\sin^2\,\theta_{ac}\,\sinh r_{ab}\,\sinh (2\,r_{bc}) + \nu_b\,\cosh^2\,r_{ab}\,\sinh^2r_{bc} \nonumber\\
&+ (\nu_a + (\nu_c-\nu_a)\,\sin^2\,\theta_{ac})\,\sinh^2r_{ab}\,\sinh^2r_{bc},\nonumber\\
\delta:=&-\frac{1}{2}\,(\nu_c - \nu_a)\,\sin (2\,\theta_{ac})\,\cosh r_{bc}\,\cosh r_{ab} + \frac{1}{2}\,(\nu_b + \nu_a + (\nu_c-\nu_a)\,\sin^2\theta_{ac})\,\sinh (2\,r_{ab})\,\sinh r_{bc},\nonumber\\
\epsilon:=&\frac{1}{2}\,\sinh(2\,r_{ab})\,\cosh r_{bc}\,(\nu_b + \nu_a+ (\nu_c-\nu_a)\,\sin^2\,\theta_{ac}) - \frac{(\nu_c - \nu_a)}{2}\,\sin(2\,\theta_{ac})\,\sinh r_{bc}\,\cosh r_{ab},\nonumber\\
\zeta:=&\frac{1}{4} (-2\,(\nu_c - \nu_a)\,\cosh(2\,r_{bc})\,\sin(2\,\theta_{ac})\,\sinh r_{ab}+ (\nu_a + 2\,\nu_b + \nu_c)\,\cosh^2r_{ab}\,\sinh(2\,r_{bc})\nonumber\\
&-(\nu_c - \nu_a)\,\cos(2\,\theta_{ac})\,(\sinh^2 r_{ab}-1)\,\sinh(2\,r_{bc})).
\end{align}
\end{widetext}

%------------------------------------------------------------------------------------------------------------------------------------------------------------------------%
\section{Covariance matrix after homodyne detection}\label{appendix:one:one}
%------------------------------------------------------------------------------------------------------------------------------------------------------------------------%
In this Appendix we derive the expression of the covariance matrix of modes $a$ and $c$ after the homodyne detection of mode $b$.

We start by adapting our vectors and covariance matrices to the notation used in \cite{spedalieri}. In particular, we need to change the basis of the vector of operators $\mathbb{X}$, and consequently the covariance matrix, from the one we employed in this work to $\mathbb{X}=(q_a,p_a,q_c,p_c,q_b,p_b)^{\text{Tp}}$, where $q_a=a+a^{\dagger}$, $p_a=-i(a-a^{\dagger})$ are the position and momentum quadratures of mode $a$ and analogous formulas hold for modes $b$ and $c$. We choose $b$ as the last mode since it is the one we intend to measure. The linear operator $\boldsymbol{K}$ which implements this change of basis is given by the (proportional to) unitary matrix
\begin{align}
\boldsymbol{K}=
\begin{pmatrix}
1& 0 & 0 & 1 & 0 & 0\\
-i & 0 & 0 & i & 0 & 0\\
0 & 0 & 1 & 0 & 0 & 1\\
0 & 0 & -i & 0 & 0 &i\\
0 & 1 & 0 & 0 &1 & 0\\
0 & -i & 0 & 0 & i & 0\\
\end{pmatrix},
\end{align}
and then the three-mode state of interest in the new basis is given by $\boldsymbol{\sigma}'=\boldsymbol{K}\,\boldsymbol{\sigma}\,\boldsymbol{K}^{\dagger}$. This has the expression
\begin{align}\label{final:three:mode:state:structurenew}
\boldsymbol{\sigma}'=2\,
\begin{pmatrix}
\alpha & 0 & \delta & 0 & \epsilon & 0\\
0 & \alpha & 0 & \delta & 0 & -\epsilon\\
\delta & 0 & \gamma & 0 & \zeta & 0\\
0 & \delta & 0 & \gamma & 0 & -\zeta\\
\epsilon & 0 & \zeta & 0 & \beta & 0\\
0 & -\epsilon & 0 & -\zeta & 0 & \beta\\
\end{pmatrix}.
\end{align}
Now, we can conveniently structure the state $\boldsymbol{\sigma}'$ matrix as
\begin{align}\label{final:three:mode:state:structurenew2}
\boldsymbol{\sigma}'=
\begin{pmatrix}
\boldsymbol{A} & \boldsymbol{C} \\
\boldsymbol{C}^T & \boldsymbol{B} \\
\end{pmatrix},
\end{align}
which will prove convenient when applying the formalism of homodyne measurement. Note that $\boldsymbol{A}$ and $\boldsymbol{B}$ would be the reduced $ac$ and $b$ states respectively, while $\boldsymbol{C}$ contains all the correlations among the $ac$ and $b$ subsystems. 
If we perform a perfect homodyne detection of a linear combination of the quadratures $q_b$, $p_b$, $x_\theta=(\cos{\theta}q_b+\sin{\theta}p_b$ the resulting two-mode (ac) state is \cite{spedalieri} 
\begin{equation}
\boldsymbol{\sigma}^{(ac)}_{{\rm out}|\theta}=\boldsymbol{A}-\boldsymbol{C}\,(\boldsymbol{\pi}_\theta\boldsymbol{B}\boldsymbol{\pi}_\theta)^{-1}\,\boldsymbol{C}^T,\label{eq:outq}
\end{equation}
where the matrices $\boldsymbol{A}$, $\boldsymbol{C}$, and $\boldsymbol{B}$, can be read out of equation \eqref{final:three:mode:state:structurenew2} and  $\boldsymbol{\pi}_\theta$ is the projector associated to a measurement of $x_{\theta}$:
\begin{equation}\label{final:three:mode:state:structurenew:B:and:pi}
\boldsymbol{\pi}_{\theta}=
\begin{pmatrix}
\cos^2{\theta} & \sin{\theta}\cos{\theta} \\
\sin{\theta}\cos{\theta}& \sin^2{\theta} \\
\end{pmatrix}.
\end{equation}
Note that $\theta=0$ and $\pi/2$ correspond to the particular cases analyzed in \cite{spedalieri}, namely the projectors of the quadratures $q_b$ and $p_b$, respectively. The inverse in Eq. (\ref{eq:outq}) needs to be understood as a Moore-Penrose pseudoinverse. As expected, the reduced ac state $\boldsymbol{A}$ is modified by the projective measurement as long as there are some correlations with mode b, which are codified in the matrix $C$. Note that for the sake of simplicity we are considering that the measurement is perfect - that is, the efficiency is 1.

As a straightforward application of equation (\ref{eq:outq}), we obtain the matrix $\boldsymbol{\sigma}^{(ac)}_{{\rm out}|\theta}$ in the notation of \cite{spedalieri}. We find 
{\small
\begin{align}
\boldsymbol{\sigma}^{(ac)}_{{\rm out}|\theta}=2\,
\begin{pmatrix}
\alpha-\frac{\epsilon^2\cos^2{\theta}}{\beta} & -\frac{\epsilon^2\sin{2\theta}}{2\beta} & \delta-\frac{\epsilon\zeta\cos^2{\theta}}{\beta} &  -\frac{\epsilon\zeta\sin{2\theta}}{2\beta} \\
-\frac{\epsilon^2\sin{2\theta}}{2\beta} & \alpha-\frac{\epsilon^2\sin^2{\theta}}{2\beta} & -\frac{\epsilon\zeta\sin{2\theta}}{2\beta} & \delta-\frac{\epsilon\zeta\sin^2{\theta}}{\beta} \\
\delta-\frac{\epsilon\zeta\cos^2{\theta}}{\beta} & -\frac{\epsilon\zeta\sin{2\theta}}{2\beta} & \gamma-\frac{\zeta^2\cos^2{\theta}}{\beta} &-\frac{\zeta^2\sin{2\theta}}{2\beta}  \\
-\frac{\epsilon\zeta\sin{2\theta}}{2\beta} & \delta-\frac{\epsilon\zeta\sin^2{\theta}}{\beta} & -\frac{\zeta^2\sin{2\theta}}{2\beta} & \gamma-\frac{\zeta^2\sin^2{\theta}}{\beta}
\end{pmatrix}, \label{eq:outqsped}
\end{align}
}
which we transform back to original basis and obtain
\begin{align}
\boldsymbol{\sigma}^{(ac)}_{{\rm out}|\theta}=
\begin{pmatrix}
\alpha-\frac{\epsilon^2}{2\beta} & \delta-\frac{\epsilon\zeta}{2\beta} & -\frac{e^{2i\theta}\epsilon^2}{2\beta} & -\frac{e^{2i\theta}\epsilon\zeta}{2\beta}\\
\delta-\frac{\epsilon\zeta}{2\beta} & \gamma-\frac{\zeta^2}{2\beta} & -\frac{e^{2i\theta}\epsilon\zeta}{2\beta} & -\frac{e^{2i\theta}\zeta^2}{2\beta} \\
-\frac{e^{-2i\theta}\epsilon^2}{2\beta}&-\frac{e^{-2i\theta}\epsilon\zeta}{2\beta} & \alpha-\frac{\epsilon^2}{2\beta} & \delta-\frac{\epsilon\zeta}{2\beta} \\
 -\frac{e^{-2i\theta}\epsilon\zeta}{2\beta}& -\frac{e^{-2i\theta}\zeta^2}{2\beta}& \delta-\frac{\epsilon\zeta}{2\beta} & \gamma-\frac{\zeta^2}{2\beta}
\end{pmatrix}. \label{eq:outqus:app}
\end{align}
A lengthy computation indicates that the resulting symplectic eigenvalues do not depend on the value of $\theta$. Moreover, the same happens with the symplectic eigenvalues of the partial transpose and the average number of photons. As a consequence, both the entanglement and the coherence are independent of $\theta$. Therefore, in the main text we  restrict ourselves to the analysis of $\boldsymbol{\sigma}^{(ac)}_{{\rm out}|q}$, which corresponds to $\theta=0$:
\begin{align}
\boldsymbol{\sigma}^{(ac)}_{{\rm out}|q}=
\begin{pmatrix}
\alpha-\frac{\epsilon^2}{2\beta} & \delta-\frac{\epsilon\zeta}{2\beta} & -\frac{\epsilon^2}{2\beta} & -\frac{\epsilon\zeta}{2\beta}\\
\delta-\frac{\epsilon\zeta}{2\beta} & \gamma-\frac{\zeta^2}{2\beta} & -\frac{\epsilon\zeta}{2\beta} & -\frac{\zeta^2}{2\beta} \\
-\frac{\epsilon^2}{2\beta}&-\frac{\epsilon\zeta}{2\beta} & \alpha-\frac{\epsilon^2}{2\beta} & \delta-\frac{\epsilon\zeta}{2\beta} \\
 -\frac{\epsilon\zeta}{2\beta}& -\frac{\zeta^2}{2\beta}& \delta-\frac{\epsilon\zeta}{2\beta} & \gamma-\frac{\zeta^2}{2\beta}
\end{pmatrix}. \label{eq:outqus:appendix}
\end{align}

%------------------------------------------------------------------------------------------------------------------------------------------------------------------------%
\section{The smallest symplectic eigenvalue of the partial transpose after homodyne detection}\label{appendix:two}
%------------------------------------------------------------------------------------------------------------------------------------------------------------------------%
We reproduce the full expression of the smallest symplectic eigenvalue $\tilde{\nu}_-^{{\rm out}|q}$ of the partial transpose of the state of modes $a$ and $c$ after homodyne detection of the modes $b$, and its counterpart for extremely close frequencies (i.e., $\nu_a=\nu_b=\nu_c\equiv\nu$). We have
\begin{widetext}
\begin{align}
(\tilde{\nu}_-^{{\rm out}|q})^2=&\alpha^2 + \gamma^2 -
 2\,\delta^2 - \frac{(\alpha\,\epsilon^2 -
  2\,\delta\,\epsilon\,\xi + \gamma\,\xi^2)}{2\,\beta}\nonumber\\
  &-\frac{1}{2\,\beta}\sqrt{-4\,\beta\,(\alpha\,\gamma - \delta^2) (\alpha\, \beta\,\gamma - \beta\,\delta^2 - \gamma\,\epsilon^2 + 2\,\delta\,\epsilon\,\xi - \alpha\,\xi^2) + (\alpha^2\,\beta + \beta\,(\gamma^2 - 2\,\delta^2) - \alpha\,\epsilon^2 + \xi\,(2\,\delta\,\epsilon - \gamma\,\xi))^2}\nonumber\\
  (\tilde{\nu}_-^{{\rm out}|q})^2=&\frac{1+2\,{\rm sh}^2_{ab}+2\,{\rm sh}^2_{bc}+10\,{\rm sh}^2_{ab}\,{\rm sh}^2_{bc}+8\,{\rm sh}^4_{ab}\,{\rm sh}^2_{ab}}{1+2\,{\rm sh}^2_{ab}+2\,{\rm sh}^2_{bc}+2\,{\rm sh}^2_{ab}\,{\rm sh}^2_{bc}}\nonumber\\
  &-4\,{\rm sh}^2_{ab}\,{\rm sh}^2_{bc}\,\frac{\sqrt{1+3\,{\rm sh}^2_{ab}+2\,{\rm sh}^2_{bc}+8\,{\rm sh}^2_{ab}\,{\rm sh}^2_{bc}+2\,{\rm sh}^4_{ab}+10\,{\rm sh}^4_{ab}\,{\rm sh}^2_{bc}+4\,{\rm sh}^6_{ab}\,{\rm sh}^2_{bc}}}{1+2\,{\rm sh}^2_{ab}+2\,{\rm sh}^2_{bc}+2\,{\rm sh}^2_{ab}\,{\rm sh}^2_{bc}}.
\end{align}
\end{widetext}

\bibliographystyle{apsrev4-1}
\bibliography{DetectorsBib}

\end{document}